\documentclass{aa}

\usepackage{graphicx}
\usepackage{txfonts}
\usepackage{amsmath}
\usepackage{booktabs}
\usepackage{natbib}
\usepackage{placeins}

\usepackage[colorlinks=true,citecolor=blue,linkcolor=blue]{hyperref}

\graphicspath{{images/}}

\providecommand{\apjl}{ApJL}
\providecommand{\aap}{A\&A}

\begin{document}

	\title{Unsupervised selection and characterisation of little red dots in JWST surveys with manifold learning}
	\titlerunning{Mapping LRDs in JWST surveys with UMAP}

	\author{Michele Ginolfi \inst{1,2}
		\and
		Filippo Mannucci \inst{2}
		\and
		Alessandro Marconi \inst{1,2}
		\and
		Francesco D’Eugenio \inst{3,4}
		\and
		Giacomo Venturi \inst{2}
		\and
		\\
		Francesco Belfiore \inst{2,5}
		\and
		Giovanni Cresci \inst{2}
		\and
		Caterina Bracci \inst{1,2,5}
		\and
		Stefano Carniani \inst{5}
		\and
		Alessandra Cozzi \inst{1}
		\and
		\\
		Roberto Maiolino \inst{3,4}
		\and
		Guido Risaliti \inst{1,2}
	}
	
	\institute{
		Dipartimento di Fisica e Astronomia, Università di Firenze, Via G. Sansone 1, I-50019, Sesto F.no (Firenze), Italy
		\and
		INAF — Osservatorio Astrofisico di Arcetri, Largo E. Fermi 5, I-50125, Florence, Italy
		\and
		Kavli Institute for Cosmology, University of Cambridge, Madingley Road, Cambridge, CB3 0HA, UK
		\and
		Cavendish Laboratory - Astrophysics Group, University of Cambridge, 19 JJ Thomson Avenue, Cambridge, CB3 0HE, UK
		\and
		European Southern Observatory, Karl-Schwarzschild-Str. 2, 85748, Garching bei München, Germany
		\and
		Scuola Normale Superiore, Piazza dei Cavalieri 7, I-56126 Pisa, Italy
	}
	
	\authorrunning{M. Ginolfi et al.}
	\date{Received XX; accepted YY}
	
	\abstract
	{
		Little red dots (LRDs) are compact, red sources discovered at high redshifts by JWST, and their physical nature and selection function remain a topic of debate. Here, we investigate whether an unsupervised machine-learning approach applied to multi-band photometry has the capacity to identify LRD-like objects, along with other populations, without relying on predefined colour-cuts.
		Using UMAP, a manifold-learning (dimensionality-reduction) method, we placed $\sim$242,000 isolated, well-measured sources from the ASTRODEEP-JWST catalogue on a 2D map, where objects with similar broadband colours, morphology, and photometric redshift lie close together.
		We then used spectroscopically confirmed LRDs to identify where LRD-like objects lie within this map. We compared the resulting areas with published colour cuts and validated our data-driven selection against archival NIRSpec spectra from the DJA.
		We find that the spectroscopically selected LRDs are concentrated in two well-defined regions with no colour-cut imposed, tracing populations that differ mainly in redshift; a difference that is imprinted on their broadband colours. The main region reaches a purity of $\simeq$0.78 at $\simeq$0.82 completeness on the spectroscopically classified subset. This result is competitive with (or cleaner than) literature colour-cuts, and yields $\sim$100 additional candidates. We also tested the method as a general tool for population discovery: the manifold recovers the locations of brown dwarfs and broad-line active galactic nuclei with no explicit criterion and isolates rare pathological outliers. Overall, unsupervised manifolds, anchored by sparse high-confidence spectroscopic labels, provide an efficient, assumption-light framework for characterising populations, comparing selection methods on a common basis, and discovering rare objects in large photometric datasets.
	}

	
	\maketitle
	
	\nolinenumbers

	\section{Introduction}
	\label{sec:intro}
	
	Among the early surprises of the James Webb Space Telescope (JWST) was the discovery of a large population of compact sources with red colours, now widely referred to as little red dots \citep[LRDs;][]{Labbe23,Matthee24,Kokorev24}.
	Their spectral energy distributions (SEDs) are characteristically 'V-shaped', with a blue rest-frame ultraviolet (UV) continuum and a steeply rising rest-frame optical continuum. A large fraction of LRDs show broad Balmer emission lines that are indicative of active galactic nuclei \citep[AGNs,][]{Greene24,Kocevski25,Matthee24}. 
	Despite rapid observational progress, the physical nature of LRDs remains debated.
	
	A clearer picture is now emerging, based on physically motivated scenarios. 
	In one such scenario, many LRDs are accreting black holes embedded in dense gas, where electron scattering (among other effects) broadens the Balmer lines and the continuum is either atrributed to thermal emission at $\sim 5000$~K or nebular emission (with some dust reddening; \citealp{Matthee26,Rusakov26,deGraaff25,Sneppen2026PaJ}). In another view, based on the `super-Eddington unification' model of \citet{MadauMaiolino26}, LRDs are the dust-obscured (type-2-like) counterparts of so-called little blue dots, where the broad lines remain visible because their obscuring column is modest (e.g. $A_V\simeq2\text{--}5$~mag instead of 20--40~mag as in type-2 AGNs). 
	Alternative interpretations have also been proposed, including dust-reddened AGNs \citep{Greene24,Kocevski25} and compact dusty stellar systems \citep{Labbe23}; 
	however, the growing spectroscopic diversity still favours a heterogeneous, possibly composite population \citep{Zhang26,PerezGonzalez26,Pan26}.
	Therefore, establishing a consistent observational definition of LRDs, followed by a clean, well-understood census and the identification of possible LRD sub-classes, is an essential prerequisite for interpreting the population.
	
	In practice, photometric LRD samples are typically assembled by combining colour-colour
	cuts, a compactness requirement, and a step designed to reject cool stars (brown
	dwarfs), whose molecular absorption mimics the LRD colours
	\citep{Kokorev24,Greene24,Barro24,Rinaldi26, Hainline26}. The specific
	thresholds, however, vary substantially from study to study: the adopted
	rest-optical `redness' threshold alone moves the recovered numbers by a factor of a few, while different choices preferentially select different parts of the underlying population. 
	As a consequence, published LRD samples overlap only partially, and the properties inferred for the population, from its number density to its broad-line (AGN) fraction and SED diversity, depend on the selection function nearly as much as on the data \citep[e.g.][]{Kocevski25, PerezGonzalez26, Barro26}.
	Spectroscopic selections \citep[e.g.][]{deGraaff25} avoid broadband-colour priors at the classification stage, but are limited to sources with high-quality spectroscopy and inherit the selection function of the parent spectroscopic sample. This is not a minor caveat: archival spectroscopic samples tend to overrepresent LRDs, because many observing programmes deliberately targeted colour-selected candidates, so a photometric prior re-enters through the target selection.

	A complementary strategy, which we pursue here, is to characterise the data
	prior to imposing thresholds, using unsupervised manifold learning
	\citep[for a recent review of unsupervised machine-learning methods in an astronomical context, see][]{Fotopoulou24}.
	The aim of this paradigm is to map complex, high-dimensional measurements into a
	low-dimensional representation (i.e. a manifold) that captures the intrinsic
	diversity of the data, without human-induced priors or restrictive empirical
	boundaries. 
	In astronomy, these methods are increasingly used to organise and classify large datasets 
	without predefined labels. For example, they have been used to compress and explore galaxy 
	spectra and SEDs \citep{Reis21, Portillo20} and to build data-driven 
	taxonomies of galaxy morphology \citep{Hocking18}. In a semi-supervised setting close to 
	the one we adopted in this work, they have also been used to classify AGNs in DESI spectra beyond 
	traditional emission-line diagnostics \citep{Alcolea26}. 
	This is precisely because structure in the data can emerge naturally rather than being imposed.
	Several dimensionality-reduction algorithms exist, including $t$-SNE
	\citep{vanderMaaten08} and a range of (variational) auto-encoders. 
	We adopted the uniform manifold approximation and projection (UMAP; \citealt{McInnes18}) 
	method for its practical advantages: 
	it scales well to large catalogues with relatively low-dimensional feature vectors, 
	tends to preserve local \emph{and} a degree of global structure, and
	demonstrates a relative level of stability with respect to its hyper-parameters. It produces an embedding, which is the low-dimensional map of the sources (i.e. the manifold introduced above, 
	realised here in two dimensions) that is well suited to both visualisation and downstream quantitative analysis. We regard the specific choice of algorithm as exploratory and return to it 
	in our assessment of robustness.
	
	Once the sources are placed on this unsupervised map, we adopt a semi-supervised strategy:
	we anchored the representation with a small number of high-confidence external labels, and then  propagate that information to the unlabelled sources nearby in the representation. 
	In detail, we used as anchors spectroscopically selected LRDs of \citealt{deGraaff25}, classified from spectral 
	shape rather than broadband colours  (although the parent spectroscopic targeting is not entirely colour-independent; 
	see Section~\ref{sec:discussion}).
	
	This step has three useful consequences. 
	First, it provides a selection that is driven by where confirmed sources 
	are actually based in the data, rather than by chosen thresholds. 
	Second, it makes the boundaries of the population, 
	and the contaminants adjacent to it, explicit and measurable. 
	Third, the same representation naturally supports the identification of outliers 
	and the discovery of rare populations, simply as regions of the manifold that are
	sparsely populated or do not resemble the bulk of the population.

	This type of data-driven perspective is timely. Recent LRD selections span two extremes: some impose increasingly restrictive criteria \citep[e.g.][]{Park26,Lin26}, at the risk of cutting away any redshift evolution of the population, while others admit X-ray- and mid-infrared (MIR)-detected sources from very large parent samples \citep[e.g.][]{Casey26,Fu26}, at the risk of including interlopers. A manifold-based representation instead shows directly whether such candidates fall within the LRD region or elsewhere.
	
	We note that, framed this way, our approach has a concrete advantage over a fully supervised
	search for LRD-like objects in the present data regime. Supervised classifiers
	require large, representative training sets, which for a rare and still
	ill-defined population do not yet exist; here a few dozen high-confidence labels
	suffice, because the unsupervised representation has already organised the data and
	the labels are needed only to indicate specifically where in that organisation the population
	lies. The same property lets us recover sources, and even entire sub-populations,
	collectively (as coherent regions of the manifold), rather than one object at
	a time.
	
	In this work we apply the approach to public JWST imaging surveys through the
	homogeneously reduced ASTRODEEP-JWST photometric catalogue \citep{Merlin24}. 
	We constructed a feature space from eight-band optical-to-MIR photometry, along with two
	morphological indicators and the photometric redshift for $\sim$242{,}000
	well-measured, isolated sources across six extragalactic fields, embedded it with
	UMAP, and anchored the embedding with the spectroscopically selected LRDs of
	\citet{deGraaff25}. We deliberately restricted the inputs to broadband photometry,
	morphology and photometric redshift as a proof of concept. The same framework
	extends naturally to spectra and resolved imaging, which we plan to consider in a future work (see Section \ref{sec:discussion}). We carefully state the assumptions and limitations of each step to allow for 
	critical assessment of the method and its results. 
	
	The paper is organised
	as follows. Section~\ref{sec:data} describes the parent sample, the features, the
	pre-processing, and the external catalogues. Section~\ref{sec:methods} presents
	the embedding, the definition of the LRD region, and the spectroscopic
	measurements. Section~\ref{sec:results} reports the structure of the manifold,
	the comparison with literature selections, the candidate sample, and the
	archival-spectroscopy results. Section~\ref{sec:discussion} offers a discussion of the
	implications of our findings, including which of the input features actually carry the LRD
	signature and the potential of the manifold to reveal other astrophysical
	populations and outliers. Finally, Section~\ref{sec:conclusions} summarises our
	conclusions. 
	\section{Data and parent sample}
	\label{sec:data}
	
	\subsection{The ASTRODEEP-JWST catalogue}
	\label{sec:astrodeep}
	We built on the ASTRODEEP-JWST photometric catalogue \citep{Merlin24}, a
	homogeneous reduction and multi-band catalogue of the principal public JWST
	extragalactic fields. It provides forced, PSF-matched aperture photometry on a
	common detection image, total-flux corrections, SExtractor morphological
	parameters, and photometric redshifts computed with EAZY (\citealp{Brammer08}), processed identically
	across fields so that colours and derived quantities are directly comparable. The
	catalogue spans seven fields: the lensing cluster Abell~2744
	\citep[GLASS-JWST and UNCOVER;][]{Treu22,Bezanson24}, CEERS in the EGS
	\citep{Bagley23,Finkelstein23}, the two JADES fields in GOODS-North and
	GOODS-South \citep{Rieke2023,Eisenstein2025,DEugenio2025,Eisenstein26}, PRIMER-COSMOS and PRIMER-UDS \citep{Dunlop21},
	and NGDEEP \citep{Bagley24}. We used the six blank (wide) fields, while excluding NGDEEP because it lacks
	coverage in the F410M medium band that we require for a uniform feature set
	(Section~\ref{sec:features}). We retained a common set of bands across all fields is
	necessary, so that position in the feature space reflects the SED shape rather than
	which filters happen to be available.
	
	\subsection{Features: Photometry, morphology, and redshift}
	\label{sec:features}

	\begin{table}[t]
		\centering
		\caption{Input features used to construct the manifold.}
		\label{tab:features}
		\small
		\setlength{\tabcolsep}{3pt}
		\begin{minipage}[t]{0.4\columnwidth}
			\centering
			\begin{tabular}{ll}
				\toprule
				Feature & Band \\
				\midrule
				\multicolumn{2}{l}{\emph{Photometry}}\\
				$\log(f_{814}/f_{356})$ & F814W \\
				$\log(f_{115}/f_{356})$ & F115W \\
				$\log(f_{150}/f_{356})$ & F150W \\
				$\log(f_{200}/f_{356})$ & F200W \\
				$\log(f_{277}/f_{356})$ & F277W \\
				$\log(f_{410}/f_{356})$ & F410M \\
				$\log(f_{444}/f_{356})$ & F444W \\
				$\log f_{356}$          & F356W (ref.) \\
				\bottomrule
			\end{tabular}
		\end{minipage}\hfill
		\begin{minipage}[t]{0.5\columnwidth}
			\centering
			\begin{tabular}{ll}
				\toprule
				Feature & Observable \\
				\midrule
				\multicolumn{2}{l}{\emph{Morphology}}\\
				\texttt{ClassStarSE}    & stellarity \\
				$\log r_{50}$           & half-light radius \\
				\midrule
				\multicolumn{2}{l}{\emph{Redshift}}\\
				$\log(1{+}z_{\rm phot})$ & photo-$z$ \\
				\bottomrule
			\end{tabular}
		\end{minipage}
		\tablefoot{The eleven features are built from the
			ASTRODEEP-JWST catalogue \citep{Merlin24}. Seven bands enter as logarithmic flux
			ratios to the F356W reference band; the F356W flux itself is retained as an
			overall-brightness feature. Morphology comes from the two SExtractor quantities
			measured on the F356W$+$F444W detection stack, and the redshift is the ASTRODEEP
			EAZY photometric redshift. All fluxes are PSF-matched aperture photometry; features
			are robust-scaled before embedding (Section~\ref{sec:preproc}).}
	\end{table}
	
	For each source, we used eight bands spanning the optical to MIR: HST/ACS F814W band (central wavelength $\lambda_{\rm c}\simeq0.81\,\mu$m) and the JWST/NIRCam bands
	F115W, F150W, F200W, F277W, F356W, and F444W (central wavelengths $1.15$, $1.50$,
	$1.99$, $2.76$, $3.57$, and $4.40\,\mu$m, respectively; broad filters with relative widths
	$\Delta\lambda/\lambda\sim0.2$--$0.3$), together with the F410M medium band
	($\lambda_{\rm c}\simeq4.08\,\mu$m, $\Delta\lambda/\lambda\sim0.1$), which helps
	isolate strong rest-frame optical emission lines from continuum at the relevant
	redshifts. This set brackets the rest-frame UV-to-optical break that
	defines the LRD SED across $z\sim3$--8.
	
	Because compactness is a defining property of LRDs, we deliberately included
	morphology in the representation through two SExtractor quantities, 
	both measured on the ASTRODEEP detection image, a stack of the F356W and F444W mosaics \citep{Merlin24}, so that compactness is assessed in the rest-frame optical where the LRD morphology is defined: the stellarity index \texttt{ClassStarSE}, which runs from 0 (resolved) to 1 (point-like), and the half-light radius \texttt{r50SE} (the radius enclosing 50\% of the flux). 
	We note that for unresolved sources, \texttt{r50SE} does not go to zero but saturates at the point spread function (PSF) half-light radius ($\sim$0.09$''$ on the F356W$+$F444W detection stack).
	
	The final feature is the photometric redshift. 
	This carries an important caveat: it assumes the photometric redshifts are reliable, whereas JWST photometric redshifts are known to include a fraction of catastrophic outliers \citep[e.g.][]{ArrabalHaro23}. Such failures displace the affected sources on the map; indeed, one of the anomalous regions we recover consists precisely of catastrophic photo-z failures.
	
	Photometry enters as colours: fluxes
	were normalised by the F356W band and log-scaled, so that the representation would
	respond to SED shape rather than apparent brightness. The reference flux in the F356W band
	itself was retained as a separate feature so that brightness information would not end up discarded.
	We adopted F356W as the reference because it is a broad NIRCam band with uniform coverage across all six fields and a high signal-to-noise ratio (S/N) for our sources, which minimises noise amplification in the colour ratios. Because the reference flux is itself retained as a feature, the choice of reference band is essentially a change of basis rather than a loss of information. These eleven features are summarised in Table~\ref{tab:features}.
	
	\subsection{Pre-processing and the parent sample}
	\label{sec:preproc}
	We required S/N$\,>2$ in each band, with one deliberate
	exception: the bluest band, HST/ACS F814W, was allowed to go undetected. A source
	that fails the threshold in F814W was retained, with its F814W flux replaced by a
	$1\sigma$ estimate from its measurement error, whereas detection was enforced in
	the seven JWST/NIRCam bands (see Section~\ref{sec:retention} for why this is required and its effect on completeness). 
	This exception keeps the sample sensitive to Lyman-break drop-outs: at $z\gtrsim5$ 
	the Lyman break  redshifts into or beyond F814W, leaving genuine high-redshift sources faint 
	or undetected there. Since F814W then probes the rest-frame UV, 
	its faintness reflects the Lyman break, not the rest-optical `redness' that defines LRDs; 
	imposing a formal F814W detection would therefore preferentially discard exactly 
	the high-redshift LRDs of interest here.
	In contrast, a detection is still required in F115W, so genuine drop-outs in that band 
	($z\gtrsim8$, where the Lyman break falls redwards of F115W) were excluded. 
	This sets an effective redshift ceiling of $z\lesssim8$: sources near this limit survive 
	only because the Lyman break lies within F115W, suppressing but not extinguishing its flux, 
	so that the permissive S/N$\,>2$ threshold still admits them. 
	The high-redshift population (median $z_{\rm phot}\simeq8.2$) we recover and discuss in Section~\ref{sec:highz}, sits precisely at this boundary; objects at higher redshift, 
	for which F115W vanishes entirely, end up falling outside the present sample.
	
	We note that the S/N$\,>2$ threshold is itself deliberately permissive. 
	Because our analysis is driven by the overall shape of the SED, 
	encoded in a learned representation, rather than by hard cuts on individual colours, 
	it is comparatively robust to the per-band noise that scatters sources across a fixed threshold, 
	and can therefore operate at a lower S/N than is otherwise typical for cut-based selections. 
	We verified that lowering the requirement further mainly admits noise-dominated
	photometry that carries little SED information, without altering any of our
	conclusions, while raising it shrinks the sample without sharpening the results; thus,
	S/N$\,>2$ is a pragmatic compromise between sample size and SED reliability.

	We additionally applied an isolation criterion, removing any source with a catalogued neighbour within $0.5''$. 
	This radius is comparable to the PSF-matched photometric aperture, the NIRCam long-wavelength PSF, and the aperture scale used in standard LRD compactness diagnostics; a closer neighbour could blend with the source and corrupt its photometry, scattering it in the feature space for reasons unrelated to its intrinsic SED.
	The isolation cut has a clear cost, in that it removes genuine sources, in particular in crowded regions and close pairs or mergers, and so introduces incompleteness that we do not attempt to correct; in return, the retained sources have cleaner, more trustworthy SEDs, while the sample remains large enough for our purposes.
	One concern deserves particular attention: because many LRDs host close star-forming companions \citep{Baggen26}, whose small separations are exactly what the isolation criterion targets, the cut could in principle remove them preferentially and bias our completeness. We tested this directly by cross-matching the \citet{Baggen26} sample, split into objects with candidate companions and solitary ones, into our sequence of selection steps.
	Within our footprint, where the sky covered by our six fields, we recover $81\%$ of the companion-hosting LRDs and $78\%$ of the solitary ones, a difference consistent with noise, and the residual $\sim$20\% loss is dominated by the S/N pre-processing and is common to both groups. 
	The reason lies in the catalogue construction rather than in the imaging resolution: 
	ASTRODEEP detects sources on a single common detection image and then measures forced, 
	PSF-matched photometry (Section~\ref{sec:astrodeep}), so a close companion that is not split off 
	as a separate entry at the detection stage is never counted as a neighbour by our $0.5''$ cut, 
	even though the native NIRCam imaging resolves finer separations.
	The isolation criterion therefore does not bias our completeness against companion-hosting LRDs.
	
	After these steps, the features were scaled with a robust (median and interquartile) scaler to limit the influence of outliers. The resulting parent sample contains 242{,}327 sources.
	
	\subsection{Completeness through the selection}
	\label{sec:retention}
	A central point for interpreting our results is that the parent sample is itself
	the product of a selection function, and we quantify its effect on the
	spectroscopic LRDs explicitly. Of the 116 unique
	spectroscopically selected LRDs of \citet{deGraaff25}, 89 ($77\%$) fall within
	the footprint and detection limits of our parent catalogue. 
	Of these, 68 ($59\%$ of the original 116) additionally satisfy the pre-processing requirements above and enter the final sample. The losses are not random in redshift: the sources removed by pre-processing have a higher median redshift ($z\simeq5.6$) than those retained ($z\simeq5.1$).
	Their redshift distributions have comparable spread (standard deviations of $1.6$ and $1.4$), so the effect is a shift to higher redshift rather than a broadening. The removed LRDs do not, however, differ in SED shape: they show the same characteristic V-shape as the retained ones, with an essentially identical rest-optical-to-rest-UV flux ratio, confirming that the loss is driven by depth rather than by a distinct spectral type.

	Waiving the detection requirement in F814W, that is, allowing it to be undetected (Section~\ref{sec:preproc}), 
	already recovers the most common dropout cases, but a detection is still required in the bluer
	NIRCam bands, so the very reddest and highest-redshift sources (faint or absent
	in F115W and F150W) continue to be preferentially removed. This residual imprint
	of the detection requirement on the high-redshift tail is the dominant bias of the
	parent selection.
	
	We therefore report completeness at two complementary levels. 
	The  intrinsic completeness measures how well a given selection recovers the
	spectroscopic LRDs that are present in our sample (relative to the 68), and
	isolates the performance of the selection itself. The \emph{end-to-end}
	completeness folds in the $\sim$59\% pre-processing retention and describes the
	fraction of all known spectroscopic LRDs recovered by the full procedure. 
	The distinction matters: the pre-processing, not the manifold selection, is the
	dominant source of incompleteness, and it is the part most readily relaxed in
	future work. 
	The natural route is to require detections only redwards of the Lyman break, rather than in every band, 
	of which the F814W exemption adopted here is a first step. 
	Extending the same logic to the NIRCam bands (e.g. by anchoring the redshift 
	on medium-band photometry such as F335M where available, as in much of the JADES footprint)
	would allow the method to follow the receding redshift frontier and, in principle, 
	admit LRD candidates based on the F115W--drop-out and the F150W--drop-out that the current detection requirement removes. 
	Doing so properly is non-trivial: our features are log-flux ratios (colours), which are undefined for an undetected band, so admitting non-detections requires reconstructing the feature space to handle upper limits and censored fluxes in a principled way, together with a field-dependent choice of which bands lie redwards of the break. We therefore left this case to a dedicated follow-up.
	
	\subsection{External catalogues and cross-matching}
	\label{sec:catalogues}
	We cross-matched the parent sample against several external catalogues, using a
	$0.4''$ positional tolerance throughout. The spectroscopically selected LRD sample
	of \citet{deGraaff25} (116 sources; selected from broken power-law fits to
	NIRSpec/PRISM continua and a compactness requirement, independent of broadband
	colour at the classification stage) provides the labelled {\it anchor} of our analysis (68 in our sample). 
	For comparison, we used two photometrically selected catalogues cross-matched into our sample, \citet{Barro26} (94 sources; colour--colour selected and spectroscopically confirmed) and \citet{Kokorev24} (187; colour--colour selected), together with the published colour criteria of \citet{Rinaldi26} and \citet{Akins25}, which we re-applied to the same parent sample (Section~\ref{sec:litmethods}).
	We also matched
	the brown-dwarf compilation of \citet{Hainline25} and \citet{Hainline26} (12 in
	our sample) and a sample of broad-line AGNs (BLAGNs; \citealp{Baccus25}). 
	Finally, we recorded the spectroscopic
	coverage from the DAWN JWST Archive (DJA; \citealp{Valentino23, Heintz24}): 14{,}982
	sources in our sample have a public NIRSpec spectrum, and 8{,}774 have a
	high-quality PRISM spectrum (\texttt{grade}$=3$, which denotes the most reliable spectra and secure redshifts; \citealp{Heintz24}).
	The latter subset is important
	because it constitutes the input over which the spectroscopic anchor selection was
	performed essentially exhaustively, which we exploit to measure purity directly. 
	\section{Methods}
	\label{sec:methods}
	
	\subsection{The embedding}
	\label{sec:embedding}
	Each source is represented by a 11D feature vector: seven
	F356W-normalised log-colours, the log reference flux, the stellarity, the log
	half-light radius, and $\log(1+z_{\rm phot})$ (Section~\ref{sec:features};  Table~\ref{tab:features}). 
	After robust-scaling the features (Section~\ref{sec:preproc}), we reduced the representation to two dimensions with the manifold-learning algorithm UMAP
	\citep{McInnes18}, using a Manhattan metric, 15 nearest neighbours, and a minimum
	separation of zero. We explored the principal hyper-parameters (the number of
	neighbours, the minimum distance, the distance metric) and several random seeds
	over a grid; the qualitative structure relevant here (the concentration of the
	spectroscopic anchor, the presence of two associated regions, and the separation
	of brown dwarfs) is recovered in all cases, and the quantitative results
	reported below do not change appreciably. We use the 2D embedding for
	discovery and visualisation: a low-dimensional view is what makes the population
	structure inspectable at a glance and supports rapid, interactive exploration of
	the manifold for discovery, which we develop into a public interactive tool in
	Section~\ref{sec:discovery}. We verify the key conclusions directly in the
	original feature space (Section~\ref{sec:featurespace}), where distances are not
	distorted by the projection.
	We adopt two dimensions for legibility and ease of presentation in print; 
	we verified that a 3D embedding recovers the same locus structure 
	and leaves our conclusions unchanged, so the choice of two dimensions is not 
	what drives our results.
	
\begin{figure*}[t]\centering
		\includegraphics[width=\textwidth]{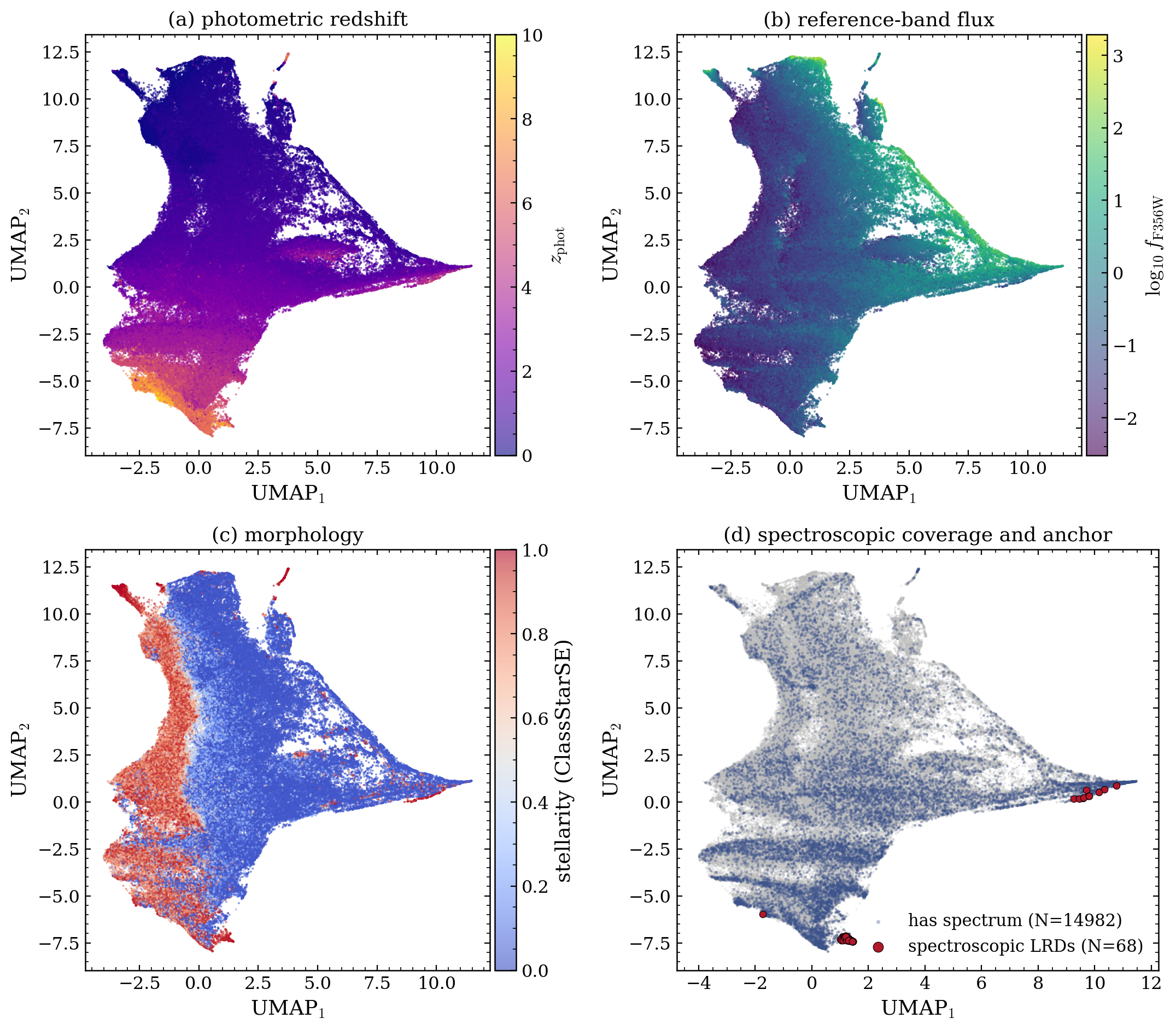}
		\caption{2D embedding colour-coded by (a) photometric redshift,
			(b) reference-band (F356W) flux, (c) stellarity, and (d) spectroscopic coverage
			with the locations of the spectroscopic LRDs. Smooth gradients in (a)--(c) show
			that the embedding encodes physical structure; panel (d) shows where the
			spectroscopic follow-up and the labelled anchor sit on the manifold.}
		\label{fig:overview}
	\end{figure*}

	The resulting manifold is not featureless: it carries clear physical structure
	(Figure~\ref{fig:overview}). Photometric redshift, reference-band flux and
	stellarity all vary smoothly across the embedding, 
	and the photometric and spectroscopic LRDs  occupy specific, well-defined regions rather than being scattered at random.
	
	\subsection{Defining the LRD region}
	\label{sec:loci}
\begin{figure*}[t]\centering
		\includegraphics[width=\textwidth]{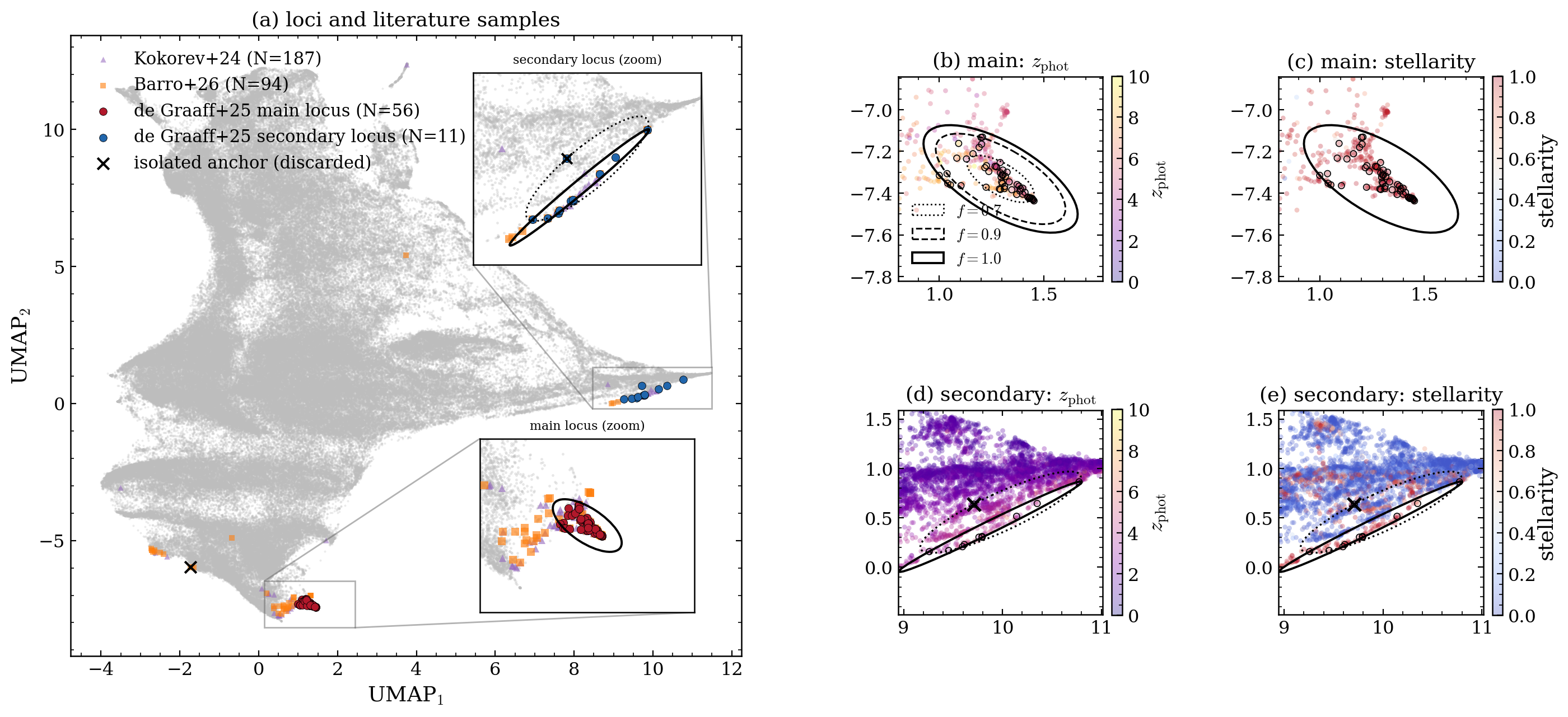}
		\caption{(a) The two anchor regions on the embedding. The spectroscopic anchors are
			shown on top (\citealt{deGraaff25} main and secondary loci); the photometrically selected
			samples of \cite{Barro26} (orange squares) and \cite{Kokorev24} (purple triangles) are overlaid
			beneath them, illustrating that the latter are more dispersed across the manifold,
			and the black cross marks the lone anchor discarded by the clustering. To keep the
			map uncluttered, the enclosing ellipses are drawn only in the two insets, one per
			locus (main, lower right; secondary, upper right): the main-region ellipse and the
			fiducial (ten-anchor) secondary-region ellipse are solid, while the full secondary
			region obtained by re-including the single bright outlier (crossed;
			Section~\ref{sec:tradeoff}) is dotted. (b--e) Zooms on each region, with the field
			colour-coded by photometric redshift and by stellarity (each panel with its own
			colourbar), showing that both loci are populated by compact, high-redshift sources.
			In panel (b) the main-region ellipse is drawn at several values of $f$, the fraction of main-island anchors it encloses (the completeness knob swept in Section~\ref{sec:tradeoff}; see Section~\ref{sec:loci}).
			In the secondary zooms (d,e) the bright outlier anchor is marked with a black cross
			and the dotted curve is the full region.}
		\label{fig:regions}
	\end{figure*}

	We located the LRD region directly from the positions of the spectroscopically 
	confirmed LRDs (the labelled anchor).
	We first grouped the
	anchor sources themselves by applying a density-based clustering (DBSCAN) to their
	standardised embedding coordinates. 
		This robustly returns two compact groups, hereafter the main and secondary regions, containing 56 and 11 anchor sources (for brevity, we also call them the main and secondary islands, using the terms islands, regions, and loci interchangeably in the following), together with a single isolated source. With only one object at hand, we cannot define an anchor region around it as we do for the two loci; so for the present analysis, we treated it as an
		outlier and set it aside. 
	We return to it in Section~\ref{sec:highz}, where its
	neighbourhood suggests a possible high-redshift extension of the main locus. The two
	groups are clearly separated on the manifold
	(Figure~\ref{fig:regions}a).
	
	We then described each group with a 2D Gaussian in the UMAP plane 
	and define an enclosing region as a Mahalanobis ellipse; namely, an iso-probability contour of that Gaussian (a distance measured in units of the local spread along each axis). 
	For the main region the size of the ellipse is set by the
	fraction, $f$, of anchor sources it encloses: scaling the Mahalanobis radius to the
	$f$-th percentile of the member distances yields a region containing a fraction
	$f$ of the anchors. 
	The completeness with respect to the anchor can thus be considered a
	design parameter that we set and vary with $f$ between 0.5 and 1.0,
	and adopting $f=1.0$, which encloses all 56 main-region anchors, as the fiducial
	value. This is meaningful precisely because the anchors are tightly clustered. The
	secondary region contains too few sources for a percentile to be meaningful, and
	is elongated, so a single Mahalanobis radius would be set by the most extreme
	source in any direction and would over-extend the long axis. We therefore enclosed
	it with the minimum-volume (L\"owner--John) ellipse, which jointly optimises the
	centre, shape ,and orientation for the smallest area containing the members. One of
	the eleven secondary anchors is a clear outlier (much brighter than the rest and
	offset from their narrow locus; Section~\ref{sec:tradeoff}), which we excluded when
	defining the fiducial secondary region. Below, we quantify the effect of re-including
	it. The zoomed-in panels of Figure~\ref{fig:regions} show that both regions are populated by
	compact, high-redshift sources, as expected for LRDs.
	
	\subsection{A comparison with literature selections}
	\label{sec:litmethods}
	Our aim here is to compare the data-driven region against the established
	photometric LRD selections in a controlled, like-for-like way. Published LRD samples are drawn
	from different catalogues, fields and depths, so their reported numbers cannot be
	compared directly with ours, or with one another. We therefore take the colour
	criteria that define each method and re-apply them to our \emph{own} parent sample,
	so that every selection acts on exactly the same objects and is judged against the
	same spectroscopic anchor. This is what allows us to place each method, together with
	our region, as a point in the common completeness--purity plane of
	Section~\ref{sec:tradeoff}.
	
	Each of these selections combines colour--colour cuts with a compactness requirement
	and a brown-dwarf rejection. To isolate the axis that actually differs between
	studies (the colour cuts), we keep the other two ingredients fixed and identical for
	all methods. For compactness, since our catalogue does not provide the small-aperture
	flux ratios used in some of these works, we adopt a single empirically calibrated
	proxy (a stellarity \texttt{ClassStarSE}$\,>0.8$, which retains 99\% of the
	spectroscopic anchor).
	One limitation of this common choice is worth noting: a relaxed colour-cut that is designed to be paired with a stricter compactness requirement (e.g.\ the unresolved-in-F444W criterion of \citealt{Rinaldi26}) is not fully reproduced by our more permissive proxy, which admits contaminants that their intended morphological cut would remove; its purity is correspondingly understated here relative to the full method.
	For brown-dwarf rejection we use F115W$-$F200W$>-0.5$, following \citet{Greene24}, which cleanly separates the known brown dwarfs in our sample.
	Any per-study magnitude or S/N floors (for instance the $\mathrm{F444W}<27$ limit of \citealt{Barro26b}) are likewise superseded by the common detection requirement of our parent sample (Section~\ref{sec:preproc}), so that every method acts on identical objects. With these common ingredients we implemented the following colour selections:
	
	\begin{itemize}
		\item \citet{Barro26b}: $\mathrm{F200W}-\mathrm{F444W}>1$ and $\mathrm{F200W}-\mathrm{F444W}>(\mathrm{F115W}-\mathrm{F200W})+0.25$, with $\mathrm{F115W}-\mathrm{F200W}>-0.5$;
		\item \citet{Kokorev24}: a two-branch cut, either $\mathrm{F115W}-\mathrm{F150W}<0.8$, $\mathrm{F200W}-\mathrm{F277W}>0.7$ and $\mathrm{F200W}-\mathrm{F356W}>1.0$, or $\mathrm{F150W}-\mathrm{F200W}<0.8$, $\mathrm{F277W}-\mathrm{F356W}>0.6$ and $\mathrm{F277W}-\mathrm{F444W}>0.7$;
		\item \citet{Rinaldi26}: $\mathrm{F150W}-\mathrm{F200W}<1.0$ and $\mathrm{F277W}-\mathrm{F444W}>0.5$;
		\item \citet{Akins25}: $\mathrm{F277W}-\mathrm{F444W}>1.5$.
	\end{itemize}
	
	\subsection{Spectroscopic measurements from archival NIRSpec data}
	\label{sec:specmethods}
	We retrieved public NIRSpec spectra from the DAWN JWST
	Archive\footnote{\url{https://s3.amazonaws.com/msaexp-nirspec/extractions/nirspec_public_v4.4.html}}
	(DJA), which provides uniformly reduced 1D extractions and an
	emission-line catalogue (\citealp{Heintz24}). Our manifold selection is purely photometric, so we used
	archival spectroscopy both to test it independently and to exploit it for
	discovery. Concretely, the spectra serve two purposes: to confirm that sources
	falling in our regions show the spectroscopic signatures expected of LRDs and
	AGNs, rather than merely their broadband colours, and to search for confirmed LRDs
	or AGNs among in-region sources that are \emph{not} part of the labelled anchor;
	that anchor was defined only over PRISM-classified spectra, leaving the sources with
	grating-only coverage as an untapped reservoir. We apply two diagnostics, each
	calibrated on the anchor as a consistency check, so that the same measurement
	applied to the anchor recovers it; both are deployed and their outcomes reported below.
	
	First, for sources with a high-quality PRISM spectrum we measure the continuum
	shape: we shift to the rest frame, convert to $f_\lambda$, and fit the power-law
	slopes in a rest-UV ($1400$--$2600\,\text{\AA}$) and a rest-optical
	($4100$--$6700\,\text{\AA}$) window. We require the characteristic V-shape
	($\beta_{\rm opt}>0$, $\beta_{\rm UV}<-0.2$, $\beta_{\rm opt}-\beta_{\rm UV}>0.5$; \citealt{deGraaff25})
	to hold in at least 95\% of Monte-Carlo realisations of the flux uncertainties, so
	that low-S/N spectra are flagged as inconclusive rather than
	spuriously selected. 
	Second, for sources with a medium-resolution grating spectrum covering H$\alpha$ we fit a linear continuum plus narrow H$\alpha$+[N~\textsc{ii}] and an optional broad H$\alpha$ component, and record a broad line when it is statistically preferred (with the broad-line fit lowering the Bayesian information criterion by $\Delta\mathrm{BIC}>10$), significant in integrated flux (S/N$~>3$), and broader than $1000~\mathrm{km~s^{-1}}$ but narrower than $7000~\mathrm{km~s^{-1}}$ (the upper bound preventing an over-broad component from absorbing the continuum and low-level noise into a spurious line).
	We restrict the
	broad-line measurement to the S/N regime in which it reliably recovers
	the anchor (Section~\ref{sec:archival}).
	\section{Results}
	\label{sec:results}
	
	\subsection{Spectroscopic little red dots concentrated on the manifold}
	\label{sec:concentration}
	Before turning to the LRDs, we recall that the manifold is physically structured
	rather than arbitrary: photometric redshift, reference-band flux and stellarity all
	vary smoothly across it (Figure~\ref{fig:overview}a--c), so the position of a source in
	the embedding already carries physically meaningful information. 
	In this context, the first and most fundamental result is that the
	spectroscopically selected LRDs are strongly localised on a manifold that was
	constructed without any knowledge of their labels (Figure~\ref{fig:overview}d). 
	We quantify the localisation in two complementary, threshold-free ways. 
	Considering, for each anchor source, its 50 nearest neighbours in UMAP space among all sources, 
	the fraction that are themselves anchors
	exceeds the value expected if the anchors were randomly distributed by a factor of
	$\sim$570; a label-permutation test rejects the null hypothesis of random
	placement at very high significance. Equivalently, a kernel-density core enclosing
	90\% of the anchors contains fewer than $10^{3}$ of the $\sim$$2.4\times10^{5}$
	sources; namely, an over-density of $\sim$230. That such a compact, over-dense region
	emerges from an unsupervised representation indicates that LRDs occupy a genuinely
	distinct part of the data space, and motivates using their location to define a
	selection. 
	
	The concentration is a property of the population, not of any single field: the manifold combines six independent JWST fields, and leaving one or two of them out leaves the spectroscopic LRDs clustered in the same region. The absolute completeness and purity shift slightly, because the set of anchors matched into the sample changes, but the overall result (that LRDs occupy a compact, over-dense locus) is unchanged.
	As noted in Section~\ref{sec:loci}, the anchors form two separated
	groups rather than one (Figure~\ref{fig:regions}); we treat the larger as the main
	island and return to the smaller in Section~\ref{sec:populations}, where we show
	that the two differ mainly in redshift while remaining distinct in their broadband
	colours (Section~\ref{sec:ablation}).
	
	\subsection{The structure is intrinsic to the feature space}
	\label{sec:featurespace}
	Because distances in a UMAP projection are not strictly metric, we verify that the
	concentration is a property of the data and not of the 2D embedding.
	Repeating the nearest-neighbour over-density measurement directly in the original
	$\sim$11D feature space (the input features themselves: flux ratios,
	reference flux, morphology and photometric redshift; see Table~\ref{tab:features}) yields a comparable
	enhancement ($\sim$520), confirming that the anchors are close to one another in
	the full representation.
	This concentration is also insensitive to the choice of reference band: rebuilding the feature space with F444W, F277W, F200W or F150W as the reference gives a comparable over-density ($\sim$500--600).
	As a complementary supervised validation, we trained a small neural-network
	classifier directly in the original 11D feature space, using five-fold
	cross-validation. This test is not used to define the LRD region, but only to ask
	whether the feature-space location of the anchors generalises to objects that were
	held out during training. The out-of-fold predictions recover $\sim$90\% of
	held-out anchor sources at a fixed threshold, and the high-probability sources
	overlap substantially with the 2D region. This confirms that the
	localisation is present in the original feature space rather than being introduced
	by the UMAP projection. The details of this test and the diagnostic probability
	distributions are given in Appendix~\ref{app:classifier}.
	
	The same feature space
	also tells the two anchor loci apart from each other: they sit somewhat closer to
	one another than to the field, yet a classifier separates their members at a
	cross-validated area under the curve (AUC) of $0.996$, dropping only to $0.991$
	when redshift is removed.
	The two loci are therefore genuinely distinct in broadband colours 
	(largely the imprint of their different redshifts, which the colours encode) 
	and only marginally closer without the explicit redshift feature; 
	we return to what distinguishes them in Section~\ref{sec:ablation}.
	
	\subsection{Completeness and purity relative to the literature}
	\label{sec:tradeoff}
\begin{figure}[t]\centering
		\includegraphics[width=\hsize]{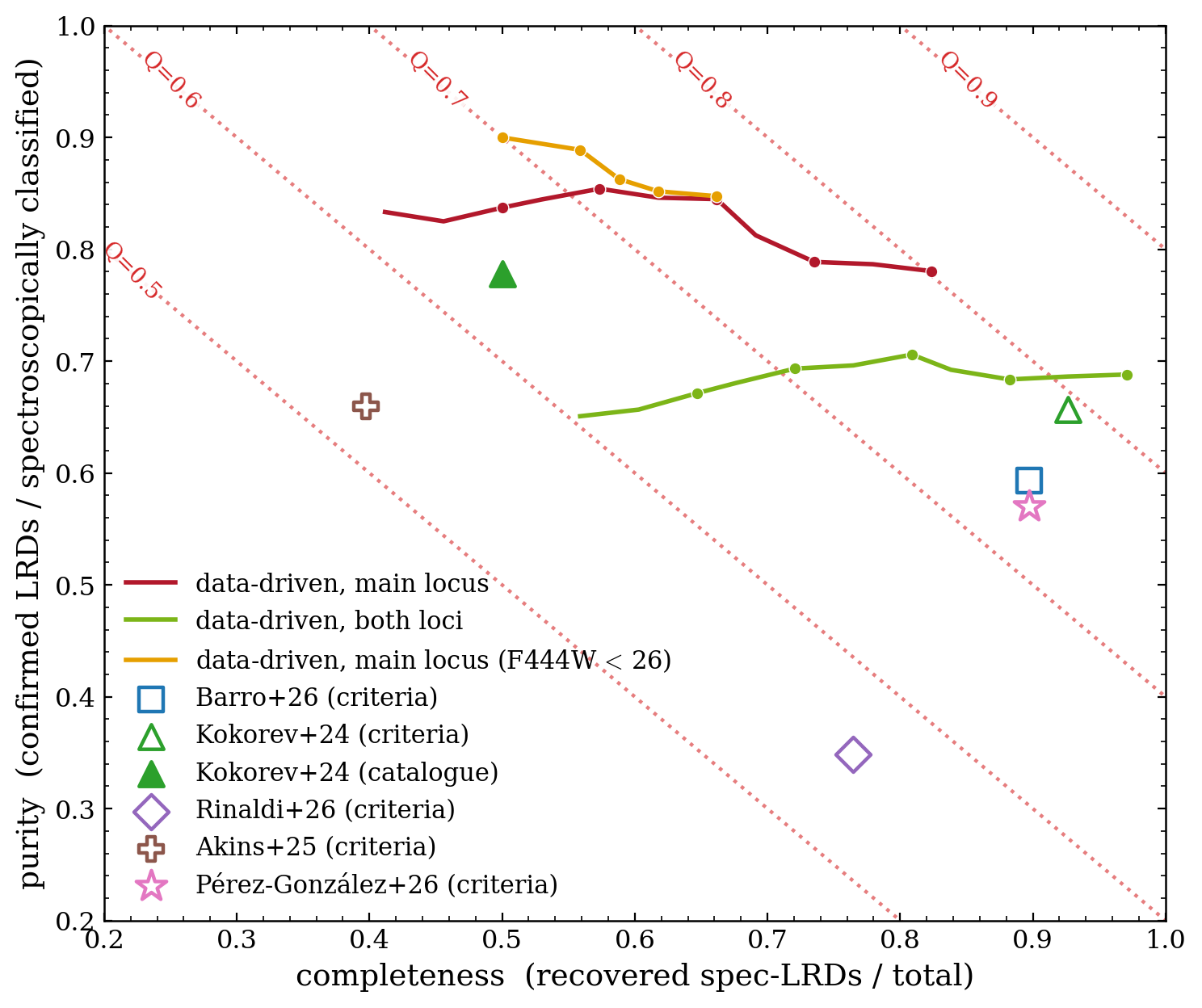}
		\caption{Completeness versus purity on the common parent sample, judged against
			the spectroscopic anchor. The data-driven selections are curves (the enclosing
			ellipse grown over a range of enclosed anchor fractions); the literature methods
			are points, as published catalogues (filled) and as their colour cuts re-applied
			to our sample (open). The lime track adds the fiducial (ten-anchor) secondary
			region to the main locus, and the orange track restricts the main locus to bright
			sources (F444W$\,<\,$26), which are purer (Section~\ref{sec:tradeoff}). Dotted red
			diagonals are lines of constant quality $Q=(\mathrm{completeness}+\mathrm{purity})/2$;
			selections towards the upper right are better in this combined sense.}
		\label{fig:tradeoff}
	\end{figure}

	We now compare the data-driven selection with the literature colour cuts in the completeness--purity plane (Figure~\ref{fig:tradeoff}, Table~\ref{tab:tradeoff}), all evaluated on the common parent sample and judged against the same spectroscopic anchor. Completeness is the fraction of anchor LRDs recovered. Purity is the fraction of confirmed LRDs among the selected sources with a high-quality (grade-3) PRISM spectrum, that is, the subset already classified in the literature from PRISM as LRD or not. A source counts as a confirmed LRD if \citet{deGraaff25} (or \citealt{Barro26}) spectroscopically classified it as one, that is, from the rest-frame continuum V-shape (a blue UV slope and a red optical slope, the $\beta_{\rm UV}$/$\beta_{\rm opt}$ criterion of Section~\ref{sec:specmethods}) rather than from broadband colours. Because \citet{deGraaff25} applied this classification essentially exhaustively to that subset, a selected, PRISM-classified source that is not a confirmed LRD is a genuine interloper rather than merely unobserved; purity here is therefore a measured quantity, not a lower limit.
	
	For the data-driven selection we trace a curve
	by varying the enclosed anchor fraction; each literature method is a single point,
	shown both as its published, positionally matched catalogue (``catalogue'') and as
	its colour cuts re-applied to our common sample (``criteria''). We omit the
	spectroscopic \citet{Barro26} catalogue point, which is $\sim$100\% pure by construction
	because its members are confirmed LRDs; the same would hold for any
	spectroscopically confirmed sample and so carries no information about the
	photometric selection.
	
	At its fiducial size, the main-region selection recovers 82\% of the in-sample
	anchors at a purity of $\sim$0.78 measured over the PRISM-classified subset (the
	selection contains 282 sources in total). We stress that this is not the purity of
	the whole photometric sample, which we cannot measure directly since most members
	lack spectroscopy; it is the confirmed-LRD fraction among the selected sources that
	carry a high-quality PRISM classification. 
	The comparison between the different photometric selections shows the expected trade-off between purity and completeness. The broadest cut of \citet{Rinaldi26} recovers $\sim0.76$ of the anchors at $\sim0.35$ purity from $\sim$2000 sources, while the reddest cut of \citet{Akins25} is purer ($\sim0.66$) but recovers only $\sim0.40$. 
	The most effective colour selections are the two-branch cut of \citet{Kokorev24} and the V-shape cut of \citet{Barro26b}, which reach high completeness ($\sim0.93$ and $\sim0.90$) at moderate purity ($\sim0.66$ and $\sim0.59$).
	We also place the selection of \citet{PerezGonzalez26}, designed for maximal completeness, on the same plane. It is drawn from the same ASTRODEEP-JWST catalogue and fields as ours, and is itself photometric: the \citet{Barro26b} V-shape cut extended by a more permissive red-colour criterion ($\mathrm{F277W}-\mathrm{F444W}>1$); their NIRSpec spectroscopy characterises the selected objects but does not enter the selection. Re-applied to our parent sample it recovers $\sim$0.90 of the anchor at $\sim$0.57 purity ($N=901$), essentially coincident with \citet{Barro26b} (Table~\ref{tab:tradeoff}, Figure~\ref{fig:tradeoff}): the more permissive criterion admits further sources but no additional spectroscopically confirmed LRDs, and these scatter across the manifold rather than forming a new concentration, while the spectroscopically confirmed LRDs it selects fall within our loci. A more inclusive colour selection is therefore captured by the same regions, at no gain in purity.
	To capture both axes at once, the dotted red diagonals in Figure~\ref{fig:tradeoff} mark lines of constant \emph{quality} $Q\equiv(\mathrm{completeness}+\mathrm{purity})/2$, a selection lying on a higher diagonal being better in this combined sense. 
	On this measure the data-driven main locus reaches the highest value, $Q\sim0.80$, marginally above the best colour selection (\citealt{Kokorev24}, $Q\sim0.79$) and well above the others ($Q\sim0.53$--$0.75$). More importantly, at a given completeness the data-driven region is the purest of all the selections, and it is a tunable curve rather than a single operating point, obtained without hand-designing any colour boundary.
	
	However, we emphasise two caveats: the region is defined using the anchor and 
	then evaluated against it,  which favours the data-driven selection, 
	and our purity is measured only over the PRISM-classified subset. 
	At the same time, the colour cuts are not a fundamentally different exercise 
	but a lower-information version of the same one, using a few filters and redshift bins 
	rather than the full SED, morphology and redshift jointly; 
	the gap in Figure~\ref{fig:tradeoff} largely reflects that difference in information content.

	Adding the secondary region raises completeness towards unity, but its photometric definition hinges on a single discrepant anchor. 
	In fact, ten of the eleven secondary anchors trace a narrow, filamentary locus;
	the eleventh, on the other hand, is a clear outlier, and a well-studied source in its own right (GN-28074; \citealt{Juodzbalis2024,Loiacono2025, Brazzini2026}). 
	It lies off this narrow axis: it is about an order of magnitude brighter than the others (a factor of $\sim$14 in the F356W normalisation flux and $\sim$11 in observed F444W, i.e.\ $\sim$2.6--2.8~mag) and at a somewhat lower redshift ($z\simeq2.3$ versus a median of $\simeq3.4$).
	A feature-by-feature decomposition of its offset from the filament shows that it is driven mainly by its observed F200W$-$F356W colour and by its brightness, not by its morphology, which is LRD-typical (it is point-like, ClassStar\,=\,1.0); the colour offset itself largely reflects the lower redshift, which shifts the rest-frame V-shape break between the bands. GN-28074 is thus still recognised as a compact, V-shape LRD, and is set aside only because, with eleven anchors, one such extreme would over-inflate the enclosing region. 
	We therefore adopt the ten-anchor filament as the fiducial secondary region, enclosed
	by its minimum-volume ellipse: 
	a thin locus (axis ratio $\sim17$) of 110 sources at $\sim$41\% purity, of which about three-quarters are compact and at anchor-like redshifts ($z\sim3.2$--$5.1$) while the remaining $\sim$25\% are resolved red interlopers.
	As discussed in Section~\ref{sec:ablation}, these are the extended sources a compactness criterion removes, though some may instead be nucleus+host systems (Saguaro-like; \citealt{Rinaldi26saguaro}) whose integrated morphology reads as resolved.
	Re-including the bright outlier (chosen objectively as the anchor whose exclusion most reduces the ellipse area) stretches the ellipse off-axis and inflates it to 698 sources, 72\% of them resolved, at only $\sim$7\% purity.
	Both versions of the region are shown in Figure~\ref{fig:regions}.
	In the completeness--purity plane (Figure~\ref{fig:tradeoff}), the ``both loci'' track is the main-locus curve with the fiducial ten-anchor secondary filament added to it. 
	On its own, the secondary locus recovers a completeness of $\simeq$0.15 at $\simeq$0.41 purity; with only $\sim$10 anchors it is a single fixed ellipse rather than a swept region, so it is one operating point rather than a curve.
	Its photometric boundary nonetheless remains only weakly constrained by the handful of available anchors, and a larger spectroscopic sample will be needed to delineate it firmly.
	
	The quoted purity is an average over the selection, but it depends on the observed
	properties of the sources, so sub-samples can be cleaner still. We find essentially
	no dependence on redshift: within the main locus the purity stays at $\sim$0.78
	from $z\simeq4$ to $z\simeq5$ and, if anything, declines slightly beyond $z\simeq6$.
	It does, however, increase with luminosity. The F444W band (the reddest broad filter,
	which samples the rest-frame optical at these redshifts, i.e.\ the red side of the
	LRD V-shape, and in which LRDs are brightest and best measured) is the natural
	luminosity proxy: restricting the main locus to F444W$\,<\,$26 raises the purity to
	$\sim$0.85--0.90 at a modest cost in completeness (orange curve in
	Figure~\ref{fig:tradeoff}), because the fainter members carry noisier photometry and
	admit more contaminants. Compactness provides a further, independent purity lever,
	which we return to in Section~\ref{sec:ablation}.
	
	\begin{table}[t]\centering\small
		\caption{Performance of literature and data-driven LRD selections.}
		\label{tab:tradeoff}
		\setlength{\tabcolsep}{4pt}
		\begin{tabular}{lccc}
			\toprule
			Selection & Compl. & Purity & $N$ \\
			\midrule
			\cite{Barro26b} criteria    & 0.90 & 0.59 & 792 \\
			\cite{PerezGonzalez26} criteria & 0.90 & 0.57 & 901 \\
			\cite{Rinaldi26} criteria  & 0.76 & 0.35 & 2027 \\
			\cite{Kokorev24} criteria  & 0.93 & 0.66 & 604 \\
			\cite{Akins25} criteria   & 0.40 & 0.66 & 139 \\
			\midrule
			Data-driven (main, $f{=}1.0$) & 0.82 & 0.78 & 282 \\
			\bottomrule
		\end{tabular}
		\tablefoot{Selections are evaluated on the common sample against the
			spectroscopic anchor, using a common compactness proxy and brown-dwarf rejection.
			Purity is measured over the PRISM-classified subset.}
	\end{table}
	
	\subsection{A sample of new candidates}
	\label{sec:candidates}
	At the fiducial size the main region contains 282 sources. Of these, 164 are
	already identified as LRDs in at least one published catalogue, with a further 11
	displaying a high-quality PRISM spectrum, but not selected by the spectroscopic
	anchor. In other words, they were examined and not classified as LRDs, so we treated
	them as interlopers and removed them. 
	They fall in the main region because their broadband colours, compactness, and photometric redshifts resemble those of LRDs, even though their PRISM spectra lack the defining features (the V-shaped continuum and/or broad Balmer lines); thus, they are likely photometric analogues rather than genuine LRDs, and it is exactly these spectroscopically identified contaminants that set the measured purity.
	
	The remaining 107 sources constitute
	a new candidate sample: they sit within the LRD region of the manifold but are
	absent from existing photometric LRD catalogues. We rank them by locus centrality,
	defined objectively as the Mahalanobis distance to the main-region centre, and
	provide the full list as a machine-readable table (see Appendix \ref{app:cand}).
	The most locus-central candidates are shown individually, as RGB cutouts and observer-frame SEDs, in Appendix~\ref{app:cand}.

	Two checks establish that this sample is not an artefact. First, it is not a
	by-product of the cross-match tolerance: the 107 candidates lie a median of
	$48''$ (minimum $3.0''$) from the nearest catalogued LRD, so none would be
	re-classified as ``known'' even if the matching radius were widened to $2''$.
	Second, although we do not impose a stringent S/N cut, the candidates
	are nonetheless well detected: all 107 have F444W S/N above 8 (median
	$\sim$56), because the all-bands requirement of Section~\ref{sec:preproc}
	already sets an effective floor. The candidate sample is also stable: resampling
	the anchor sources (bootstrap) leaves the candidate set largely unchanged, with a
	mean Jaccard overlap of $0.94$ (the size of the intersection divided by the size of the union of the two candidate sets) and 103 of 107 candidates recovered in at least
	80\% of resamples.
	
\begin{figure*}[t]
		\sidecaption
		\includegraphics[width=12cm]{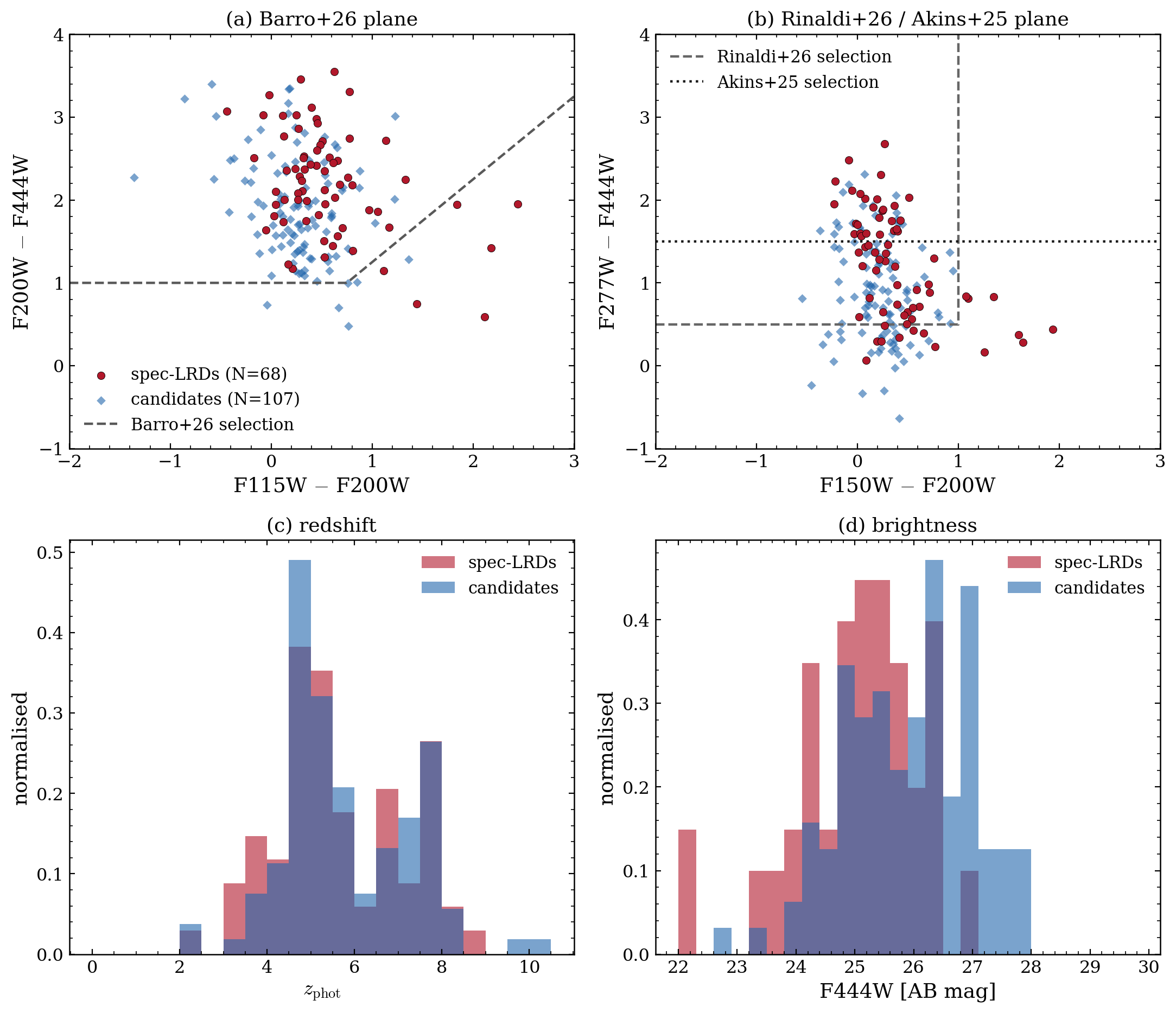}
		\caption{New candidates (blue) compared with the spectroscopic LRDs (red) in
			the two colour--colour planes used by the literature, with the corresponding selection boundaries overplotted, and in photometric redshift
			and F444W magnitude. The candidates are compact and at LRD-like redshifts but are
			bluer in the rest-optical, below the strict redness thresholds.}
		\label{fig:candidates}
	\end{figure*}

	The diagnostic breakdown of why the colour cuts miss these sources is informative (Figure~\ref{fig:candidates}). The candidates are compact (only 3\% fail the compactness proxy) and largely consistent with the more inclusive colour selections: only 6\% fall outside the \citet{Barro26b} V-shaped region, and 28\% fail the F277W$-$F444W$ ~ >0.5$ threshold of \citet{Rinaldi26}. What excludes them is the strictest redness cut: 81\% fail the F277W$-$F444W$~>1.5$ threshold of \citet{Akins25}. In other words, the candidates are compact and at LRD-like redshifts (see Section~\ref{sec:populations}), but with somewhat bluer optical colours and slightly fainter F444W magnitudes than the reddest catalogued LRDs. 
	This is consistent with the view that the strictest colour cuts sample only the reddest extreme of the LRD population, and shows concretely which sources are left out.
	
	That many candidates satisfy the more inclusive colour cuts yet are absent from the published catalogues shows that their colours are not exceptional: the same colour selection yields a different sample when applied to a different reduction, depth and compactness measure (see Section~\ref{sec:discussion}).
	For \citet{Barro26b}, which spans all our fields, footprint plays no role and depth only a minor one (only a minority of candidates fall below its depth, $\mathrm{F444W}>27$); most candidates are colour-consistent, compact and bright, and are missed instead through differences in the compactness measure (\citealt{Barro26b} use an aperture flux ratio, $\mathrm{F444W}[0.5'']/\mathrm{F444W}[0.2'']<1.5$, that our stellarity proxy does not reproduce) and in the photometry, which we take from the independent ASTRODEEP reduction \citep{Merlin24}.
	
\begin{figure*}[t]
		\sidecaption
		\includegraphics[width=12cm]{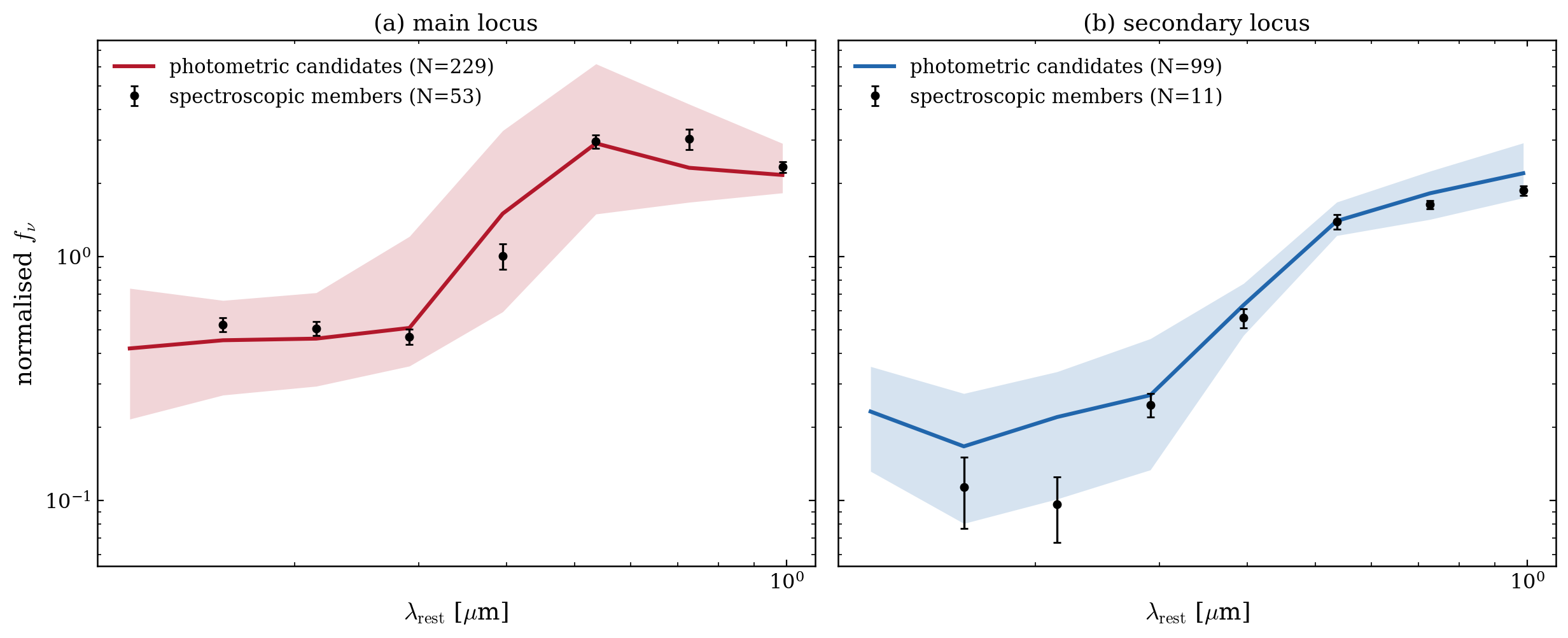}
		\caption{Rest-frame stacked SED of the two loci (left: main; right: secondary, the
			fiducial ten-anchor filament). In each panel the photometric candidates (coloured
			line: median with 16--84th percentile band) are compared with the
			spectroscopically confirmed members (black points, with error bars giving the
			standard error of the binned median). Each source is normalised to its median band
			flux so the stack reflects SED \emph{shape}, and fluxes are stacked as $f_\nu$ by
			binning in rest-frame wavelength. Both loci show the characteristic LRD V-shape, and
			in each the photometric candidates track the spectroscopic members; the secondary
			spectroscopic stack is noisy because it contains only 11 objects.}
		\label{fig:avgsed}
	\end{figure*}

	In line with these findings, the candidates also display the broadband SED shape expected for LRDs.
	Figure~\ref{fig:avgsed} stacks the rest-frame SEDs of the region members,
	separating those with a spectroscopic redshift from the purely photometric
	candidates. In both loci the photometric candidates reproduce the characteristic
	LRD V-shape traced by the spectroscopically confirmed members (a blue
	rest-UV, a minimum near the rest-frame optical break, and a red
	rest-optical rise), confirming that they share the SED of the confirmed population
	rather than being contaminants. Such agreement is expected by construction, since
	both are selected as neighbours on the same manifold, but it is a useful
	consistency check.
	
	\subsection{Brown dwarfs separate without an explicit cut}
	\label{sec:bd}
\begin{figure*}[t]\centering
		\includegraphics[width=\textwidth]{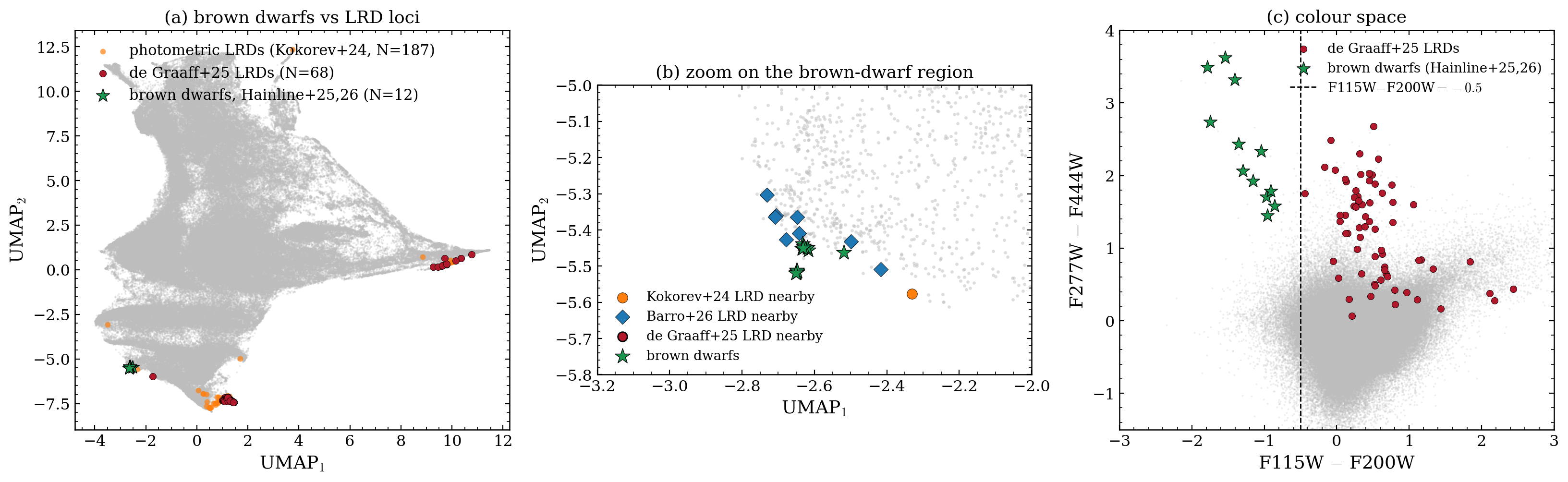}
		\caption{(a) Brown dwarfs (green) relative to the LRD loci. (b) Zoom on the
			brown-dwarf region, with nearby photometric LRD candidates highlighted as
			possible contaminants. (c) The same in colour space; brown dwarfs are blue in
			F115W$-$F200W.}
		\label{fig:bd}
	\end{figure*}

	No brown-dwarf rejection was applied when constructing the manifold, yet the known
	brown dwarfs occupy a region well away from the LRD loci (Figure~\ref{fig:bd}):
	none fall within the fiducial main region, and their median Mahalanobis distance
	to the main-region centre is $\sim$32, compared with $\sim$1.2 for the LRDs.
	The separation is driven by the blue F115W$-$F200W colours of the cool stars,
	which the manifold encodes naturally; the data-driven approach therefore
	reproduces, without being told to, the standard colour rejection. The zoomed view
	of the brown-dwarf region (Figure~\ref{fig:bd}b) shows that a small number of
	photometrically selected LRD candidates from the literature lie nearby; these are
	plausible stellar contaminants of purely photometric selections, and their
	location adjacent to, but outside, our LRD region illustrates how the
	representation can flag such ambiguous cases.
	
	\subsection{Two regions, two populations}
	\label{sec:populations}
\begin{figure*}[t]\centering
		\includegraphics[width=\textwidth]{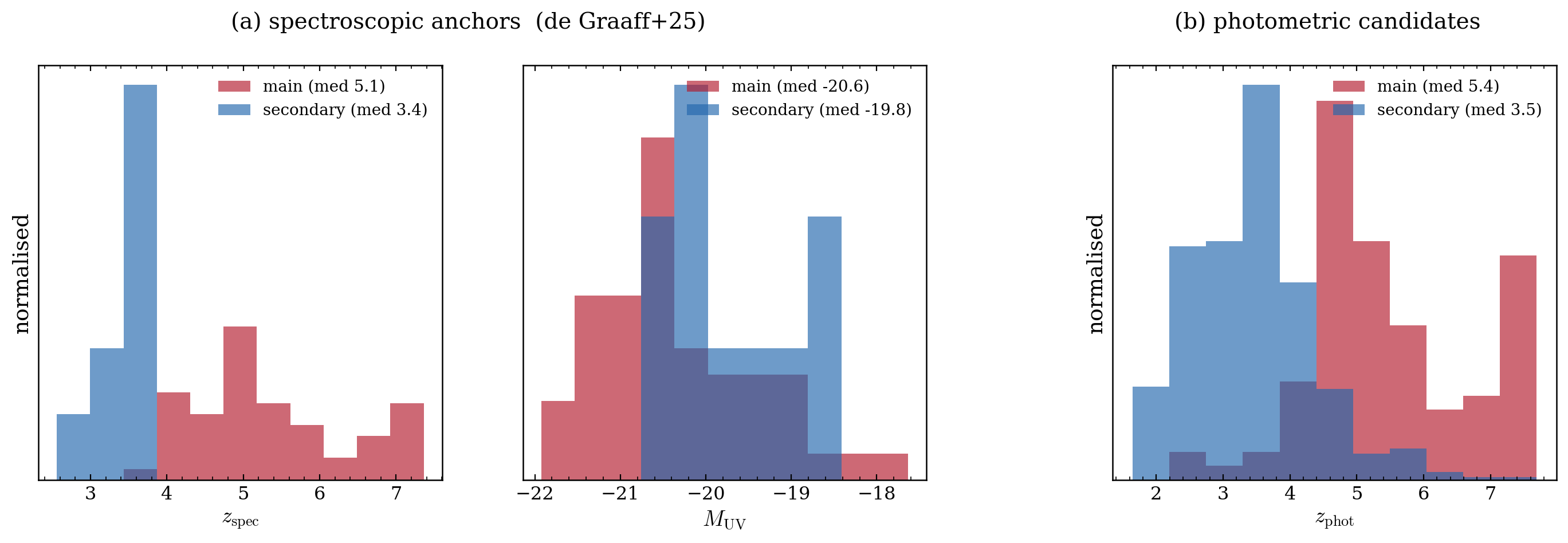}
		\caption{Two loci are two populations that differ mainly in redshift and
			rest-UV luminosity. (a) Spectroscopic anchors, from the public
			\citet{deGraaff25} fits: the secondary island (blue) lies at lower spectroscopic
			redshift and is fainter in $M_{\rm UV}$ than the main island (red). (b) Photometric
			candidates: their photometric redshifts follow the same two-population split,
			indicating they are drawn from the same populations as the anchors. The remaining
			\citet{deGraaff25} fitted quantities, which show no separation between the loci, are given in Appendix~\ref{app:populations}.}
		\label{fig:twopop}
	\end{figure*}

	The two anchor islands are not redundant, but their main apparent difference is
	redshift. Drawing on the publicly available spectroscopic measurements from the
	\citet{deGraaff25} catalogue, the clearest differences are in redshift (Figure~\ref{fig:twopop}a): 
	the main-island anchors have a median spectroscopic redshift of $5.1$ and the secondary-island
	anchors $3.4$, and the main island is the more luminous in the rest-UV
	(brighter $M_{\rm UV}$). 
	The remaining fitted quantities (UV slope,  Balmer-break strength, characteristic continuum temperature,
	Balmer decrement, and the H$\alpha$ and [O\,\textsc{iii}]\,$\lambda5007$ equivalent widths) 
	show no clear separation between the two (Appendix~\ref{app:populations}); 
	the distinction is thus driven by redshift and luminosity more than by
	emission-line properties. This is consistent with the colours-only analysis of
	Section~\ref{sec:ablation}: removing redshift and morphology brings the two loci
	only marginally closer and they remain separable in colour, but because broadband
	colours encode redshift, much of that colour difference reflects the redshift offset
	itself, and in the rest frame both loci show the same characteristic LRD V-shape
	(Figure~\ref{fig:avgsed}).
	The photometric candidates inherit the same split, with median photometric
	redshifts of $5.4$ and $3.5$ in the main and secondary islands respectively
	(Figure~\ref{fig:twopop}b), indicating that the candidates are drawn from the
	same populations as the anchors rather than being a separate contaminant set. Both
	loci show the characteristic LRD V-shape in the rest frame (Figure~\ref{fig:avgsed}),
	the secondary at systematically lower redshift; within each locus the photometric
	candidates and the spectroscopically confirmed members trace the same shape, though
	the secondary spectroscopic stack is noisy given its 11 members.

	\subsection{Archival spectroscopy: Consistency check and new AGNs}
	\label{sec:archival}
	Finally, we use archival NIRSpec spectra to both verify our picture and look for
	confirmed LRDs and AGNs among in-region sources beyond the anchor. The relevant
	sources are those that lie in one of the two regions (main and secondary) and have a spectrum but no high-quality PRISM classification (the anchor having classified the PRISM-observed sources
	already); these are predominantly sources with only medium-resolution grating
	data.
	
	The continuum diagnostic (i.e. the V-shaped continuum fit to the low-resolution PRISM spectra; Section~\ref{sec:specmethods}) confirms our framework but
	yields no new continuum-classified LRDs: applied to the anchor it recovers the great majority of
	sources, but among the in-region sources without a PRISM classification it returns
	no confident detections, simply because almost none of them have the PRISM
	continuum data the test requires. This is the expected outcome: over sources
	with adequate continuum spectra, the anchor selection is already essentially
	complete.
	
\begin{figure*}[t]\centering
		\includegraphics[width=\textwidth]{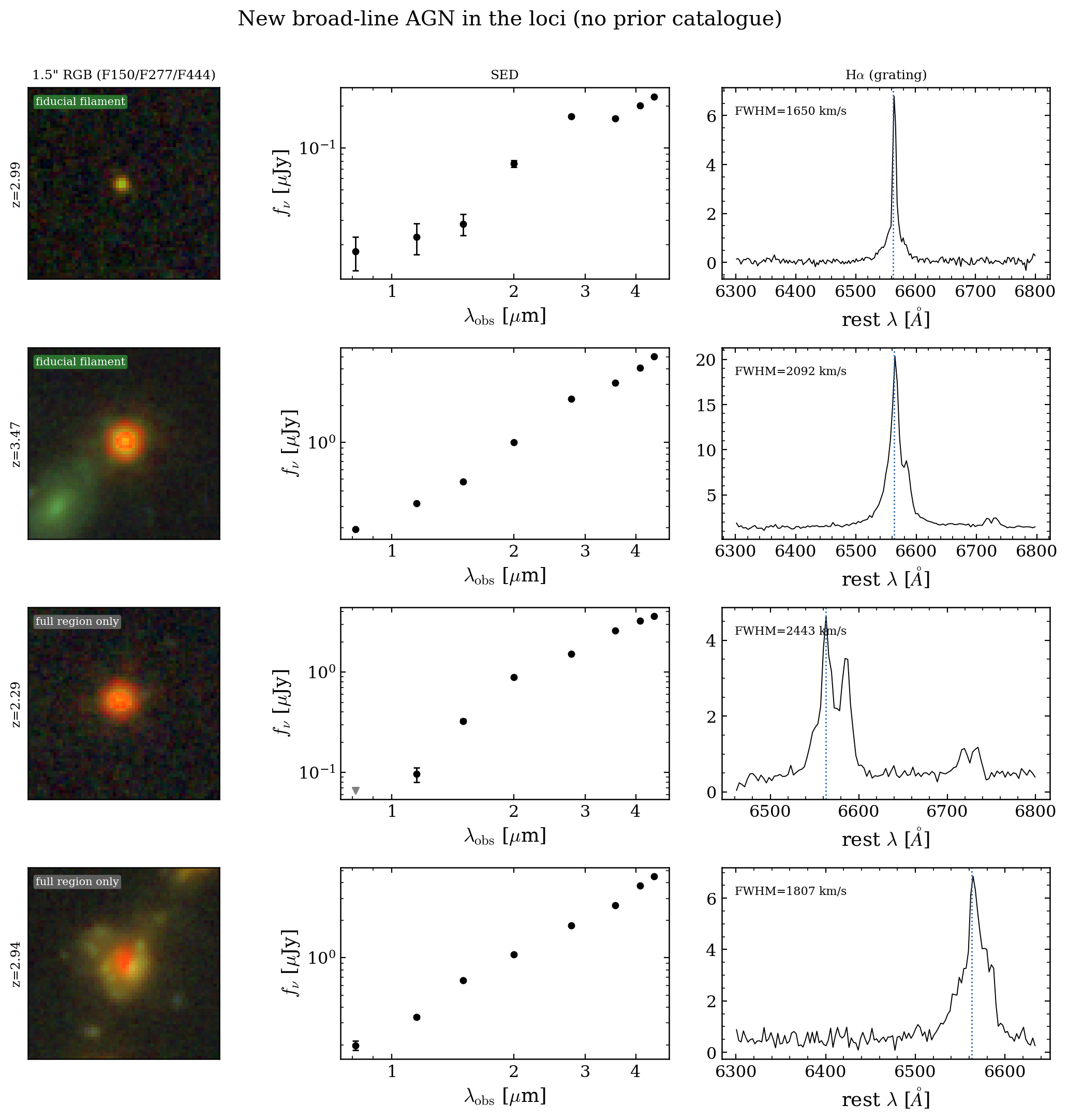}
		\caption{Four new BLAGNs with no prior catalogue identification
			(Section~\ref{sec:archival}). Top two panels: Two objects that also fall
			within the fiducial ten-anchor secondary region (the outlier-excluded filament; see Figure~\ref{fig:regions}). The lower two lie in the full secondary region only, as labelled in the left-most panels.
			For each target, we report a $1.5''$ RGB cutout (left), the observer-frame SED with uncertainties (centre), and the rest-frame grating spectrum around H$\alpha$ (right).}
		\label{fig:newagn}
	\end{figure*}

	The broad-H$\alpha$ diagnostic (i.e. a fit for a broad H$\alpha$ component in the medium-resolution grating spectra; Section~\ref{sec:specmethods}) is more productive. As a calibration we apply it to
	the anchor sources that have grating coverage of H$\alpha$: the recovery of a
	broad line is set by the H$\alpha$ S/N, reaching $\sim$93\% (13 of
	14) above a S/N of $\sim$15, comparable to the broad-line fractions
	reported for spectroscopic LRD samples; the lower recovery at low S/N values
	reflects the data, not the method. We then apply the same fit to the in-region
	sources that have grating coverage of H$\alpha$ but no PRISM classification. As the
	lower-redshift, H$\alpha$-accessible sources concentrate in the secondary region, we search within this whole region, including the area contributed by the bright outlier anchor
	(Section~\ref{sec:tradeoff}). 
	Of the 55 such grating sources, 22 have H$\alpha$ in
	the reliable regime (H$\alpha$ S/N$~\gtrsim15$), and of these 11 show a statistically preferred
	broad component. After removing marginal and floor-limited fits and any source
	within $1''$ of an anchor, nine are robust broad-line detections:
	four are BLAGNs with no prior catalogue
	identification (broad H$\alpha$ widths of $1.6$--$2.4\times10^{3}\,\mathrm{km\,
		s^{-1}}$), and the remaining five are spectroscopic confirmations of
	photometrically selected candidates. All four of the new AGNs lie in the
	secondary, lower-redshift region ($z\simeq2.3$--3.5); two of the four also fall
	within the fiducial ten-anchor filament, the other two in the full region only.
	Each missed the anchor
	selection because it lacks a high-quality PRISM spectrum, and missed the
	photometric catalogues because it falls outside their colour or footprint
	selection. 
	Their cutouts, SEDs and H$\alpha$ profiles are shown in Figure~\ref{fig:newagn}.
	We present these as broad-line active nuclei rather than LRDs: they are identified from broad H$\alpha$ rather than the V-shape continuum, some are extended or bluer than classic LRDs, and, lacking a grade-3 PRISM classification, they do not enter our purity measurement (Section~\ref{sec:tradeoff}). They populate the lower-purity secondary region, consistent with its more heterogeneous nature.

	\section{Discussion}
	\label{sec:discussion}
	
	\subsection{Little red dots: Selection and populations}
	\label{sec:two-confirm}
	
	The recurring difficulty in assembling LRD samples is that the selection function
	shapes the inferred population. The colour cuts in current use differ mainly in
	the adopted rest-optical redness threshold, and our diagnostic breakdown
	(Section~\ref{sec:candidates}) shows directly what this implies: the sources that
	the strict cuts omit are not failures of compactness or S/N, but
	compact, high-redshift objects whose optical colours are simply not red enough to
	pass. The data-driven region recovers many of these because it is defined by where
	spectroscopically confirmed LRDs actually lie in the data, not by a colour
	boundary. We are deliberately cautious about how far to push this comparison. Our
	completeness is defined relative to the spectroscopic anchor, which is itself a
	bright, V-shape-selected sample, and our purity is measured over the subset with
	high-quality PRISM spectra. 
	Since the anchor is used both to define the region and to judge it, these in-sample numbers are optimistic; a fairer test is the cross-validation in Section~\ref{sec:featurespace}, where some anchors are held out during training and the method is scored on how well it recovers them.
	
	Even with these caveats, the outcome is encouraging: using photometry, morphology and photometric
	redshift alone, the data-driven selection is already competitive with established
	colour cuts, and because the dominant source of incompleteness is the pre-processing
	(Section~\ref{sec:retention}) rather than the manifold selection itself, there is
	clear room to push it further, for instance by relaxing the detection requirement at
	the bluest wavelengths or by adding spectroscopic and imaging features.
	
	The choice of anchor also matters. 
	We use the \citet{deGraaff25} sample because its classification is spectroscopic and independent of broadband colours, so it does not import a broadband-colour prior at the classification stage into a method whose purpose is to avoid one.
	One clarification is in order. The \citet{deGraaff25} selection is not line-based: the low-resolution PRISM cannot resolve broad H$\alpha$, so their classification rests on the continuum V-shape (the UV and optical slopes; Section~\ref{sec:specmethods}). It is therefore a spectroscopic, higher-resolution measurement of the SED shape rather than an emission-line classification, that is, a refined, continuum-based counterpart of a colour selection rather than one orthogonal to it. When we call it independent of broadband colours, we mean it is independent of the photometric colours we embedded $-$ not of the SED shape in general. 
	
	The \citet{deGraaff25} selection is also distinct from a spectroscopically confirmed, but photometrically
	selected sample, such as that of \citet{Barro26}: although its members are bona
	fide LRDs, only about half of them fall in the main region, against more than
	four-fifths of the \citet{deGraaff25} sample. This is apparent in
	Figure~\ref{fig:regions}a, where the \citet{Barro26} and \citet{Kokorev24} photometric samples, overlaid on
	the manifold, are visibly more dispersed than the tightly clustered spectroscopic anchors. 
	The difference is informative (it reflects the broader, fuzzier reach of a colour-based selection), but it also means that using such a sample to anchor the region would re-import colour-selection effects and blur the locus. We therefore keep it as a comparison and confirmation set rather than as an anchor.
	We stress that the residual targeting bias of spectroscopic samples is a caveat on the anchor's completeness, not on the method itself: the LRD locus is a structure present in the data, which the anchor \emph{locates} rather than creates (Sections~\ref{sec:concentration},~\ref{sec:featurespace}). A colour-biased anchor would change which known LRDs populate it (an effect we quantify as the completeness relative to the \citealp{deGraaff25} sample; Section~\ref{sec:retention}), but not the existence or position of the locus. The \citet{deGraaff25} sample remains the least broadband-colour-biased set available, and hence the natural anchor.

	Finally, the two islands are not merely an artefact of the method but a real feature of the data (Section~\ref{sec:ablation}): 
	the manifold separates the spectroscopic LRDs into two regions differing mainly in redshift 
	and rest-UV luminosity (Section~\ref{sec:populations}),
	sharing the same characteristic V-shape in the rest frame (see Figure \ref{fig:avgsed}).
	Notably, the only new spectroscopic confirmations we obtain (four broad-line active nuclei) 
	all lie in the lower-redshift, secondary region, in part because H$\alpha$ is accessible there, 
	marking it as a promising target for dedicated follow-up. 
	Whether the two regions are physically distinct populations or the same population 
	seen across a range of redshifts remains open: 
	the spectroscopic stacks of the anchors are indistinguishable in continuum and line properties beyond the redshift shift (Appendix~\ref{app:populations}), and their full physical characterisation (the relative contributions of active nuclei and host light, dust content and emission-line properties) is beyond the present scope and is deferred to future work.
	
	\subsection{A possible high-redshift extension of the main locus}
	\label{sec:highz}
\begin{figure*}[t]\centering
		\includegraphics[width=\textwidth]{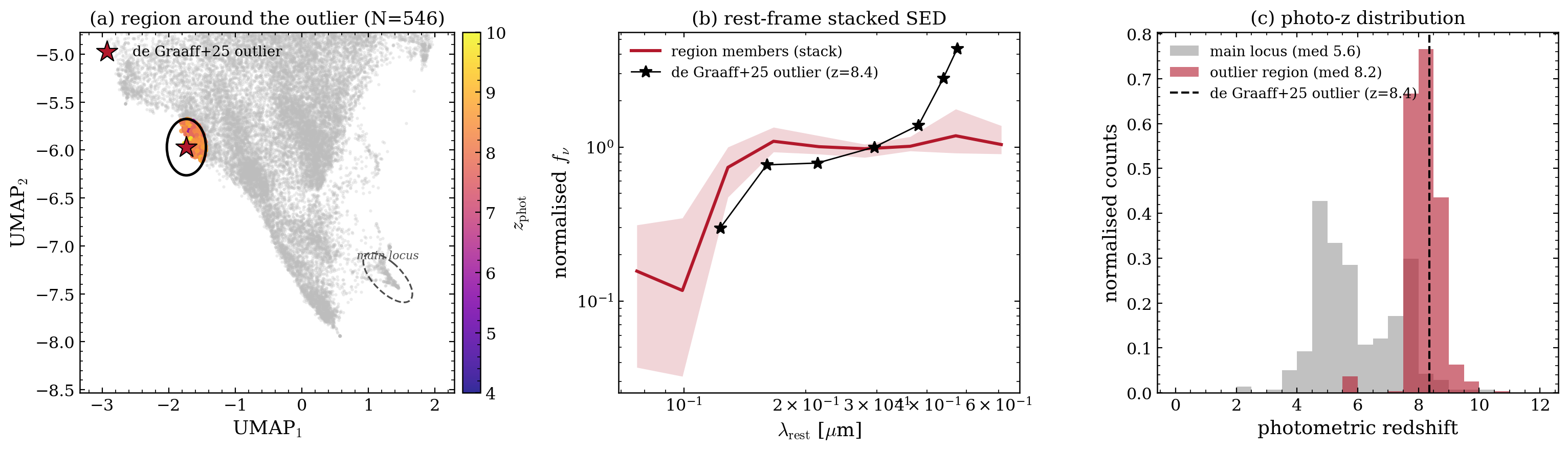}
		\caption{Discarded \cite{deGraaff25} outlier and its neighbourhood (speculative; a single
			spectroscopic redshift). (a) UMAP zoom: Field colour-coded by photometric
			redshift, the outlier (star), the assumed circular region (mean semi-axis of the
			main-locus $f{=}1$ ellipse), and the main locus (dashed). (b) Binned rest-frame
			stacked SED of the region members (normalised to the median band flux) with the
			outlier's own SED overlaid. (c) Photometric-redshift distribution of the region
			against the main locus; the region peaks near $z\simeq8$.}
		\label{fig:highz}
	\end{figure*}

		We now return (albeit speculatively) to the single \cite{deGraaff25} anchor that our
		clustering set aside as an outlier (Section~\ref{sec:loci}). 
	It sits well away from
	both loci, yet its neighbourhood is not empty: it is occupied by a coherent
	population with a distinctive signature, a large flux gap
	between the two bluest bands (a median F814W$-$F115W of $2.0$~mag, against $0.3$ in
	the main locus), the classic imprint of a Lyman break falling between them. Because
	a single object does not permit the anchor-based region-drawing used for the two
	loci, we adopt for this preliminary look a deliberately simple proxy, and we stress
	it as a caveat: a circular region centred on the object, with a radius equal
	to the mean of the semi-minor and semi-major axes of the main-locus $f{=}1$ ellipse.
	This region contains 546 sources whose photometric redshifts pile up at very high
	values (median $z_{\rm phot}\simeq8.2$, with 98\% at $z>7$), far above the main locus members
	(median $\simeq5.6$); the anchor object itself lies at $z\simeq8.4$
	(Figure~\ref{fig:highz}). 
	Their rest-frame stacked SED shows an LRD-like red rest-optical continuum, though the anchor object itself is redder than the stack and the rest-frame coverage at $z\sim8$ is too limited for a clean V-shape measurement (Figure~\ref{fig:highz}b).
	Taken together, this
	suggests that the region is a plausible higher-redshift continuation of the main
	LRD locus, resolved by the manifold as a separate concentration simply because the
	Lyman break has moved redwards through the filter set. We emphasise the obvious
	limitation: the interpretation rests on a single spectroscopic redshift and on
	photometric redshifts that are themselves uncertain at $z\sim8$; confirming it, and
	establishing whether these are bona fide $z\sim8$ LRDs, will require dedicated
	spectroscopy. Independently of the anchor, the region is interesting in its own
	right: it coincides with the highest-redshift part of the manifold in the
	photometric-redshift map (top-left panel of Figure~\ref{fig:overview}). A natural
	follow-up is spectroscopic confirmation of the most promising members, prioritised by
	proximity to the \citet{deGraaff25} object and by compactness; in parallel, in a companion
	study we will mine the archival spectroscopy already available for these sources for
	LRD features, noting that at $z\simeq8$ H$\alpha$ is redshifted out of the NIRSpec
	range, so the search must rely on other diagnostics (rest-UV lines, the
	Balmer break, or [O\,\textsc{iii}] and H$\beta$ where covered). We flagged this region as
	a concrete, testable prediction of the method.
	
	\subsection{Which features carry the LRD signature}
	\label{sec:ablation}
\begin{figure*}[t]\centering
		\includegraphics[width=\textwidth]{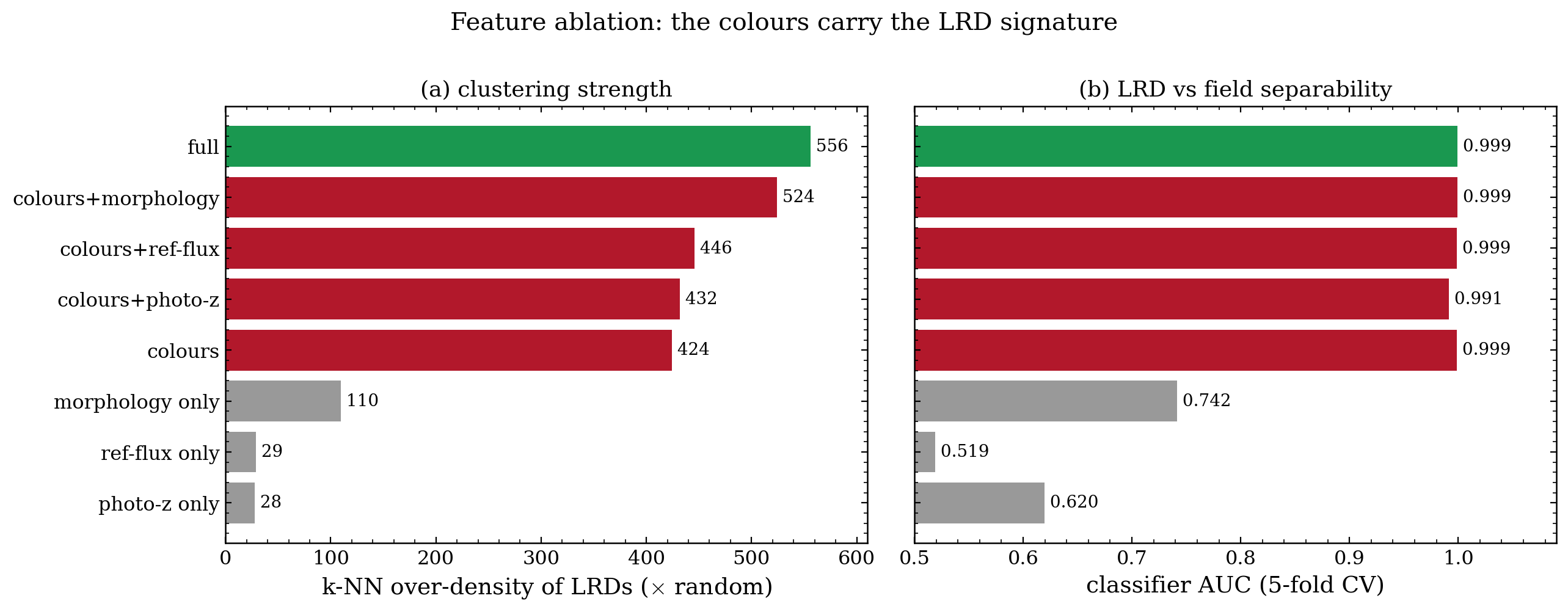}
		\caption{Feature ablation. For each subset of the input features, the
			feature-space $k$-nearest-neighbour over-density of the spectroscopic LRDs (left)
			and the cross-validated classifier separability (right). The broadband colours
			alone carry the LRD signature; morphology and photometric redshift add little and
			are weak on their own.}
		\label{fig:ablation}
	\end{figure*}

	Every photometric LRD selection in the literature treats compactness as an
	essential ingredient, alongside the colour cuts. The representation lets us ask a
	sharper question regarding which of the input features actually carry the information that
	identifies an LRD. We address it with a feature-ablation experiment, measuring the
	clustering of the spectroscopic anchor for different subsets of the features (grouped as in Table~\ref{tab:features}), both
	in the feature space directly (a $k$-nearest-neighbour over-density, independent of
	any embedding) and through a cross-validated classifier
	(Figure~\ref{fig:ablation}). The result is unambiguous: the seven broadband
	\emph{colours} alone reproduce essentially the full localisation of the LRDs (an
	over-density of $\sim$420 against the random baseline, and a classifier
	AUC of $0.999$), whereas morphology or photometric redshift on their
	own are weak discriminators (AUC $0.74$ and $0.62$). Adding
	morphology, reference flux or photometric redshift on top of the colours changes
	the clustering only marginally. In other words, the photometric fingerprint of the
	LRDs is carried by the SED shape, and the additional features are largely
	redundant for the purposes of identification.
	
\begin{figure*}[t]\centering
		\includegraphics[width=\textwidth]{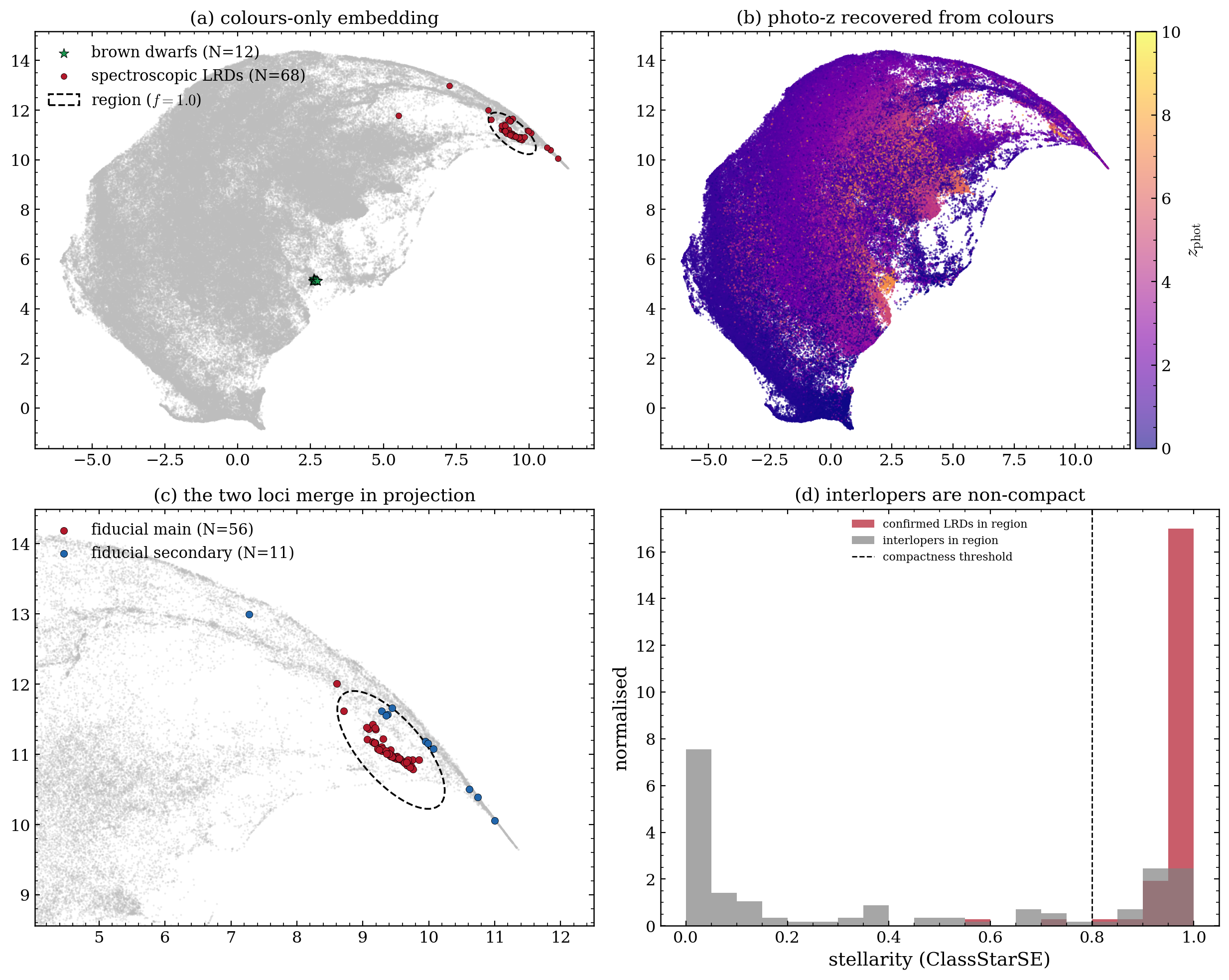}
		\caption{Colours-only embedding (morphology, reference flux and photometric
			redshift switched off). (a) Spectroscopic LRDs still form a single
			concentration and the brown dwarfs remain separate. (b) Photometric redshift is
			recovered as a smooth gradient. (c) Colours-only projection where the two fiducial
			populations merge into one concentration, although they remain separable in the
			underlying feature space (Section~\ref{sec:ablation}). (d) Interlopers admitted into the colours-only region, which are predominantly non-compact, i.e.\ these are exactly the sources a compactness criterion removes.}
		\label{fig:nomorph}
	\end{figure*}

	We can confirm this directly by rebuilding the embedding from the colour ratios alone,
	switching off morphology, reference flux and photometric redshift
	(Figure~\ref{fig:nomorph}). The spectroscopic LRDs again form a single, compact,
	over-dense concentration (Figure~\ref{fig:nomorph}a), with a $k$-nearest-neighbour
	over-density comparable to the fiducial run. Three further points follow. 
	
	First,
	the photometric redshift is effectively re-learned from the colours: the
	colours-only manifold shows a clear redshift gradient
	(Figure~\ref{fig:nomorph}b), a simple regressor predicts $z_{\rm phot}$ from the
	two coordinates with a coefficient of determination $R^{2}\simeq0.3$ (the 2D coordinates alone capture about a third of the redshift variance, modest but non-trivial for a 2D embedding built without redshift), and, most tellingly, supplying redshift as an explicit
	feature barely alters the LRD clustering, consistent with colours encoding
	redshift. 
	
	Second, the colours-only map clarifies the relationship between the two
	fiducial populations. They are no longer laid out as two separate islands but merge
	into a single connected concentration (Figure~\ref{fig:nomorph}c); this merging,
	however, is largely a property of the projection. In the underlying feature space
	the two loci stay distinct even without redshift: a classifier separates their
	members from the broadband colours alone at a cross-validated AUC of $\sim$0.99,
	against $\sim$0.996 with all features, that is, only marginally closer. The two
	populations are thus genuinely different in their observed colours, not merely a
	photometric-redshift artefact. Because broadband colours nonetheless encode redshift
	(the colours-only map recovers $z_{\rm phot}$ as a smooth gradient,
	Figure~\ref{fig:nomorph}b), much of this colour difference reflects the redshift
	offset between the two, and in the rest frame both show the same characteristic LRD
	V-shape (Section~\ref{sec:populations}); the lower-redshift locus is the one hosting
	the broad-line nuclei of Section~\ref{sec:archival}.
	
	Third, and importantly, morphology is not useless: its role is purity,
	rather than identification. To compare like with like, we define the colours-only
	region exactly as in the main analysis (Section~\ref{sec:loci}): we cluster the
	anchors, enclose each group with an ellipse containing all of its members
	($f=1.0$), and measure purity over the same spectroscopically classified subset.
	At this fixed, identical region-drawing choice, the colours-only region is markedly
	less pure than the fiducial one (a purity of $\sim$0.4 against $\sim$0.8), and the additional
	interlopers are predominantly resolved sources: $\sim$70\% of them fall below our
	compactness threshold (Figure~\ref{fig:nomorph}d), and applying that threshold
	\emph{a posteriori} restores the purity to $\sim$0.7 at essentially unchanged
	completeness. Compactness, the staple of photometric LRD selection, therefore acts
	in this framework as a filter against extended red interlopers, not as part of the
	core identification. 
	One caveat here is worth investigating in its own right: not all
	of these resolved interlopers need to be genuine contaminants. 
	An object whose light is a compact red nucleus embedded in a more extended host can have its integrated morphology (and hence \texttt{ClassStarSE}, which is measured on the F356W$+$F444W detection stack and therefore in the rest-frame optical; Section~\ref{sec:features}) biased towards `resolved' when the host's rest-optical emission is extended. Its SED can nonetheless stay nucleus-dominated, especially if the photometry is extracted within an aperture that excludes much of the host, so the source still lands among the LRDs on the manifold.
	Establishing how many of the colours-only interlopers are of this kind, by
	inspecting their cutouts individually or with dedicated morphological modelling, is
	left to future work.
	We emphasise the natural caveats: first, the anchor is itself
	selected to be compact, so the colours that identify it are the colours of compact
	red sources; second, the parent sample is moreover restricted to isolated, well-measured
	objects. Within these limits, the experiment shows that, given confirmed LRDs,
	their broadband colours alone carry the photometric signature, and points towards
	an even simpler, colour-only data-driven selection in which morphology enters only
	as a purity control.
	
	\subsection{The method as a general discovery accelerator}
	\label{sec:discovery}
	Beyond the specific case of LRDs, the manifold itself is a general tool for population discovery. 
	Two modes are particularly useful. First, any external
	sample can be placed on the existing manifold by position, so that one can ask
	immediately where a new class of objects falls, which known sources it resembles,
	and whether it forms a coherent region. Second, the same representation makes
	rare objects, sparsely populated regions, outliers and ``bridges'' between
	populations apparent, without any prior definition of what to look for (see e.g. \citealp{Reis21}). 
	As a proof of concept we have made the embedding interactive through a web-based
	visualiser\footnote{\url{https://micginolfi.github.io/compressedUniverse/}}. It
	displays the manifold and, for any selected source, fetches the imaging cutout and
	the NIRSpec spectrum from the DAWN JWST Archive (\citealp{Valentino23, Heintz24}) 
	and shows the SED in both the
	observed and rest frames together with colour--colour diagrams; selecting a region
	returns the stacked SED and the distributions of the enclosed sources. Such
	interactivity turns the representation into an exploratory instrument for
	identifying and vetting populations directly.
	
\begin{figure}[t]\centering
		\includegraphics[width=\hsize]{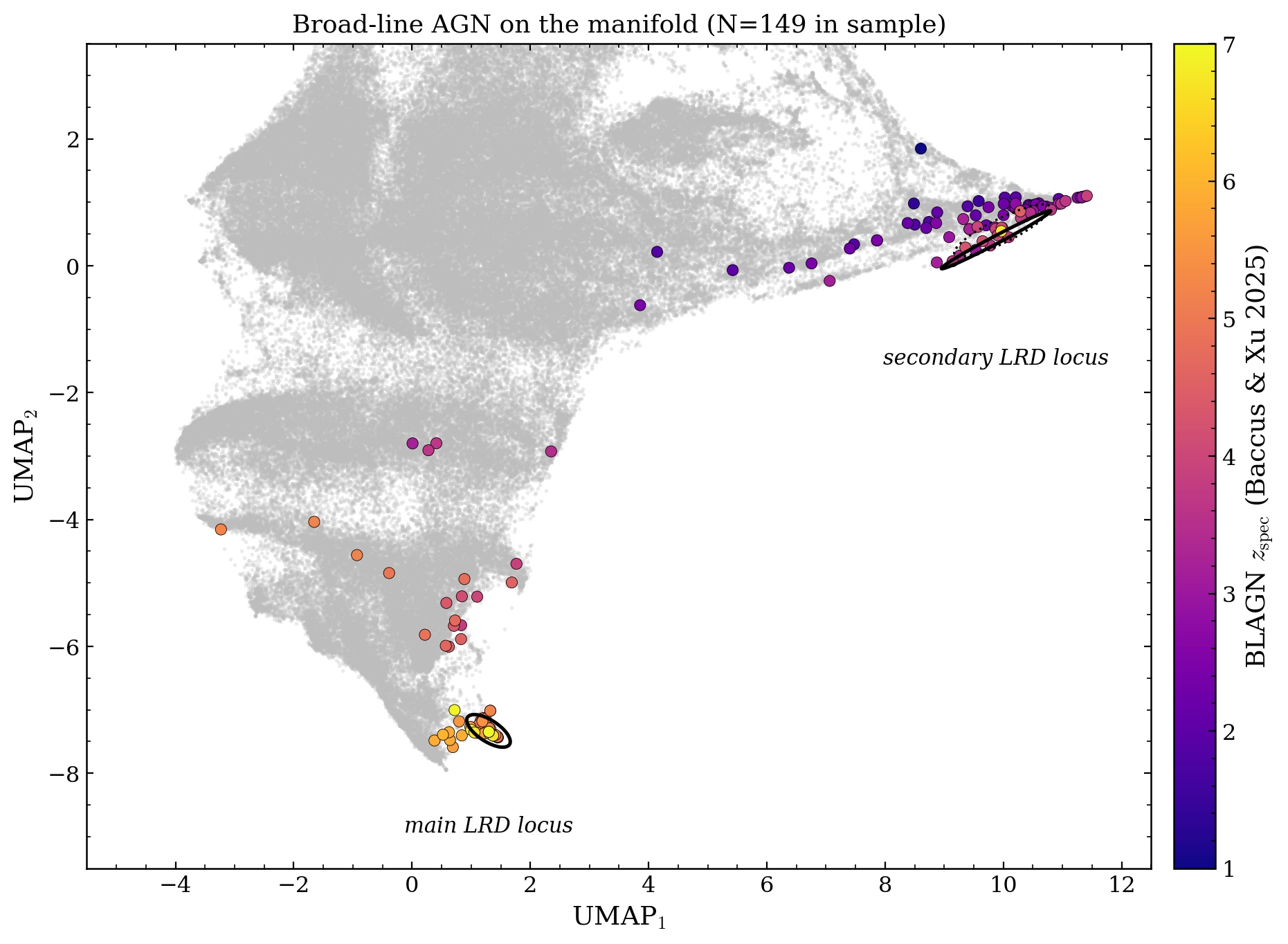}
		\caption{BLAGNs from the \citet{Baccus25} JWST/NIRSpec census located on the
			manifold (the 149 of 252 that fall in our parent sample), colour-coded by
			spectroscopic redshift, with the main and secondary LRD loci ellipses for reference
			(the dotted curve is the full secondary region). The view is zoomed on the lower
			part of the plane, where the BLAGNs and the loci lie. The BLAGNs are not randomly scattered: they concentrate inside and around the LRD loci and in a few extreme regions,
			and follow the same redshift trend as the loci (high redshift near the main locus,
			low redshift near the secondary).}
		\label{fig:blagn}
	\end{figure}

	Known populations placed on the manifold fall where their physical properties would
	predict, not at random. The brown dwarfs are one example, already described
	(Section~\ref{sec:bd}) and with no rejection step, they occupy a region well away from
	the LRD loci; BLAGNs) are another example. Projecting the
	JWST/NIRSpec BLAGN census of \citet{Baccus25} onto the map
	(Figure~\ref{fig:blagn}), 149 of the 252 sources fall within our parent sample, and
	they are far from uniformly distributed: about 40\% (58) land in the two LRD loci
	and most of the rest concentrate around them and in a few extreme regions of the
	plane (e.g. the high-flux corner at the right-hand end). Colour-coding by
	redshift reveals the same trend seen for the LRDs themselves: the higher-redshift
	BLAGNs gather around the main locus and the lower-redshift ones around the secondary
	locus. The overlap with the LRDs is substantial but partial: 26 of the 68
	spectroscopic LRDs are also BLAGNs, while more than half of the BLAGNs lie
	outside the loci, consistent with the LRDs being one particularly compact corner of
	the wider accreting-black-hole population. This is exactly the question the
	representation answers at a glance: where a new sample sits, what it resembles, and
	how coherent it is.
	
	A concrete demonstration is that the same representation that localises the LRDs
	also isolates other coherent
	groups of sources: regions of the manifold occupied by objects that share a
	distinctive signature. Exploring the embedding interactively with
	this viewer, we find several such regions; here we highlight
	three that illustrate the outlier- and population-finding capability, and that
	double as a useful data-quality diagnostic. In each case the grouping arises
	because the sources share a common distortion of their measured SED, which the
	unsupervised representation captures without being told what to look for.
	
\begin{figure*}[t]\centering
		\includegraphics[width=\textwidth]{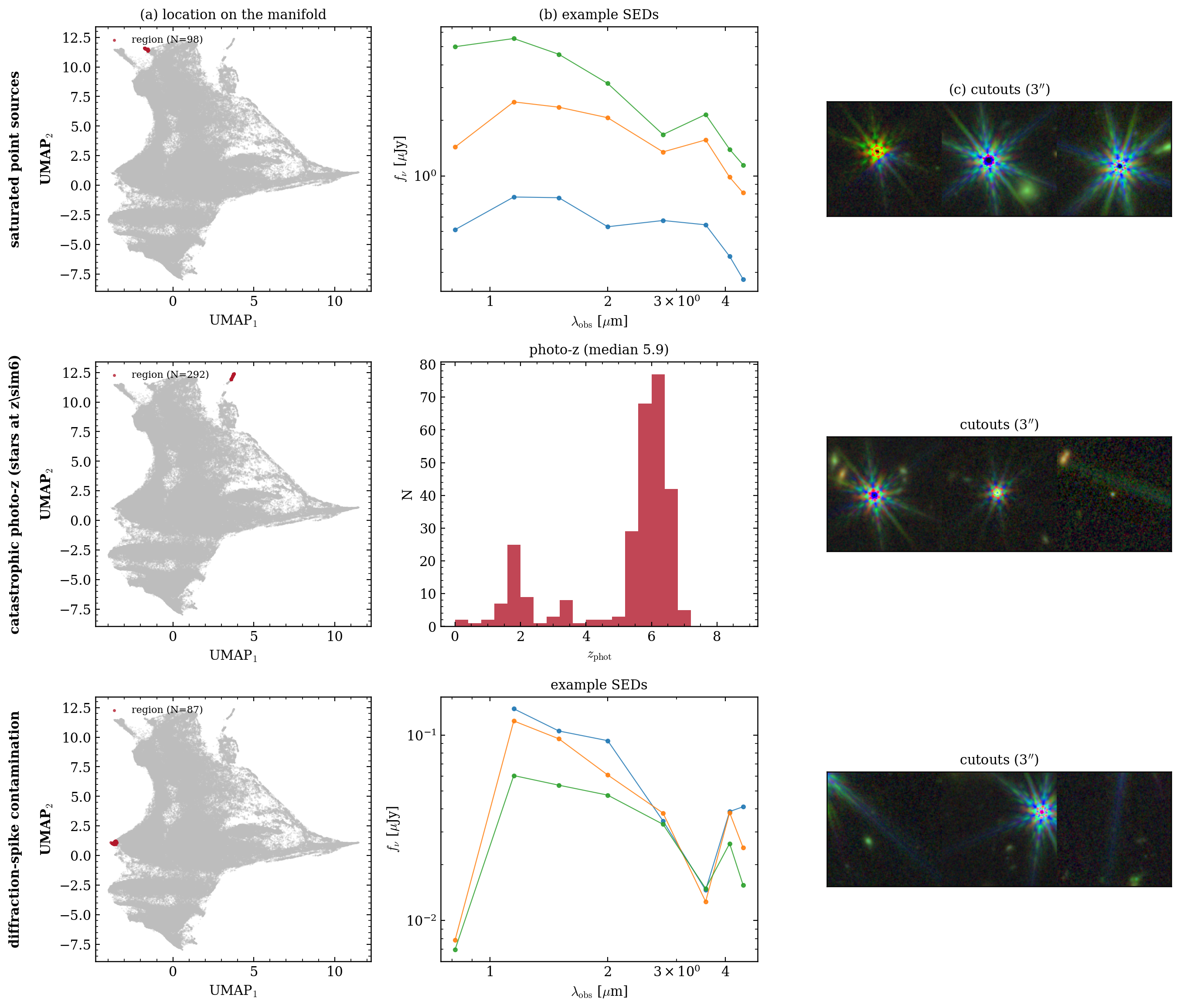}
		\caption{Three peculiar regions that the manifold isolates without any prior
			definition, one per row. \emph{Top}: saturated bright stars. \emph{Middle}:
			Catastrophic photometric-redshift failures (Galactic stars fitted with
			high-redshift galaxy templates, so their catalogue $z_{\rm phot}$ piles up near
			$z\simeq6$). \emph{Bottom}: Sources contaminated by the diffraction spikes of a
			nearby bright star. Columns: (a) Region's location on the embedding. (b) Example
			SEDs of indicative members, or, for the middle row, the region's
			photometric-redshift distribution. (c) $3''$ RGB cutouts of indicative members. Each
			region is a coherent group sharing a common, distinctive distortion of the measured
			SED, so it forms its own concentration on the manifold and is surfaced as an
			unsupervised data-quality diagnostic.}
		\label{fig:outliers}
	\end{figure*}

	The first region (Figure~\ref{fig:outliers}, top row) is a compact clump of bright point
	sources (stellarity $\simeq1$, photometric redshift $\simeq0$). Their cutouts show
	saturated cores and strong diffraction spikes: these are bright, saturated stars,
	whose photometry is corrupted in a characteristic way that places them together on
	the manifold. The second region (Figure~\ref{fig:outliers}, middle row) is also dominated by
	point sources whose imaging is plainly stellar, yet their catalogue photometric
	redshifts pile up at $z\simeq6$ (median $5.9$, with $\sim$80\% assigned $z>4$).
	These are catastrophic photometric-redshift failures: Galactic stars fitted by
	high-redshift galaxy templates. Because the (erroneous) redshift is one of the
	input features, it displaces them into the high-redshift part of the map, but
	their stellar SEDs make them a coherent, anomalous island that stands out from the
	genuine high-redshift sources around it: the representation effectively flags the
	failure. The third region (Figure~\ref{fig:outliers}, bottom row) is a clump of faint
	sources whose cutouts reveal that they lie along the diffraction spikes of a
	nearby bright star; the spike contaminates their photometry and imprints a shared,
	non-monotonic SED distortion that again groups them together.
	
	None of these groupings was sought in advance and none of them required any threshold:
	they emerge because the embedding organises sources by the shape of their measured
	SED, so any population with a common spectral signature (astrophysical or
	instrumental) forms its own region. This has immediate practical value as an
	unsupervised quality-control step, automatically surfacing saturated sources,
	imaging artefacts and ancillary-catalogue failures that would otherwise
	contaminate a sample. More importantly, it is the same mechanism that enables
	genuine discovery: a previously unknown class of objects, or an outlier unlike the
	bulk of the population, will likewise occupy a distinct or sparsely populated
	region. Combined with the interactive viewer, which allows arbitrary external
	samples to be projected onto the manifold by position and any region to be
	inspected through its stacked SED, cutouts and distributions, this makes the
	representation a practical instrument for both vetting and discovery. A systematic
	census of the additional astrophysical populations the manifold reveals is beyond
	the scope of this paper and is deferred to future work.
	
	\section{Conclusions}
	\label{sec:conclusions}

	In this work, we explore an unsupervised, data-driven route to identifying and
	characterising LRDs in JWST surveys, embedding $\sim$242{,}000
	well-measured, isolated ASTRODEEP-JWST sources \citep{Merlin24} in a
	low-dimensional manifold of broadband colours, morphology, and photometric redshift
	(Section~\ref{sec:embedding}), anchoring it with the spectroscopically selected
	LRDs of \citet{deGraaff25}. Our main findings and their limitations are as follows:

	\begin{itemize}
		\item LRDs occupy a distinct region of the manifold
		(Section~\ref{sec:concentration}). Without any colour-cut, the spectroscopic LRDs
		end up concentrated into a compact region that is over-dense by more than two orders of
		magnitude, split into two associated sub-regions (Section~\ref{sec:loci}).
		
		\item The structure is intrinsic, not a projection artefact
		(Section~\ref{sec:featurespace}). The same localisation is present in the original
		feature space, and a cross-validated classifier recovers $\sim$90\% of held-out
		spectroscopic LRDs, confirming that the location generalises.
		
		\item Colours carry the LRD signature (Section~\ref{sec:ablation}). A
		feature-ablation experiment shows that the seven broadband colours alone reproduce
		essentially the full localisation of the LRDs (classifier AUC $0.999$), while the
		morphology and photometric redshift are individually weak discriminators
		(AUC $0.74$ and $0.62$) and largely redundant for identification once the colours
		are included; the morphology nonetheless improves the purity.
		
		\item The selection is competitive and interpretable
		(Section~\ref{sec:tradeoff}). When defining the region with Mahalanobis ellipses, whose
		size sets the completeness with respect to the anchor, the main region reaches
		$\sim$0.82 completeness at a directly measured purity of $\sim$0.78 over the
		PRISM-classified subset (not of the whole photometric sample), placing it at the high-purity end of the completeness-purity plane, at completeness comparable to the best literature colour cuts (Table~\ref{tab:tradeoff}). For comparison, on the same sample the best hand-tuned colour selections reach high completeness, but at a lower level of purity (\citealt{Kokorev24}: 0.93 completeness, 0.66 purity; \citealt{Barro26b}: 0.90 completeness, 0.59 purity); the data-driven region achieves the highest purity at comparable completeness and the highest combined quality $Q$, without any hand-designed colour boundary.
		
		\item A new candidate sample is provided (Section~\ref{sec:candidates}). The region
		yields 107 compact, high-redshift candidates absent from existing catalogues,
		robust to anchor resampling and to the cross-match tolerance. The strict literature redness cuts miss them because their rest-optical colours are not red enough; those that satisfy the more inclusive cuts are absent from the published catalogues due to differences in the compactness measure and in the photometry $-$ and not in colour (see Section~\ref{sec:candidates}).
		
		\item Contaminants separate naturally (Section~\ref{sec:bd}). Brown dwarfs
		fall well outside the LRD region with no explicit rejection step, while the
		representation flags ambiguous, adjacent photometric candidates.
		
		\item Two populations emerge from the manifold (Section~\ref{sec:populations}). The two regions correspond to two populations differing mainly in redshift, which shifts the V-shape break through the filter set and with it their broadband colours, a distinction inherited by the photometric candidates.

		\item  A predicted higher-redshift extension is apparent (Section~\ref{sec:highz}). The
		single anchor discarded as an outlier sits in a coherent neighbourhood of 546
		sources whose photometric redshifts pile up at $z_{\rm phot}\simeq8$ and 
		whose stacked rest-frame SED shows an LRD-like red continuum, 
		suggesting a higher-redshift
		continuation of the main locus resolved separately only because the Lyman break has
		moved redwards through the filter set. This is a concrete, spectroscopically testable
		prediction, and rests for now on a single spectroscopic redshift.
		
		\item A set of new BLAGNs are confirmed (Section~\ref{sec:archival}). A broad-H$\alpha$
		analysis of archival NIRSpec spectra (Section~\ref{sec:specmethods}), calibrated on
		the anchor ($\sim$93\% recovery at an adequate S/N), confirms four
		BLAGNs with no prior catalogue identification, all in the
		lower-redshift secondary region, plus five confirmations of photometric candidates.
		
		\item  Our manifold approach can also serve as a tool for discovery beyond LRDs (Section~\ref{sec:discovery}). The same representation isolates other coherent populations and outliers as distinct
		regions of the manifold, including a group traceable to catastrophic
		photometric-redshift failures (Section~\ref{sec:discovery}), illustrating the method's
		broader use as a discovery instrument. 
		
	\end{itemize}
	
	These results come with clear caveats. Our completeness is defined relative to a
	bright, spectroscopically selected anchor, and the purity is measured only over the
	subset of sources with high-quality (grade-3) PRISM spectra (Section~\ref{sec:tradeoff}). 
	The dominant incompleteness is set by the pre-processing, and in particular by the requirement of a detection down to F115W, the bluest NIRCam band (only HST/ACS F814W is exempt), which imposes an effective redshift ceiling of $z\lesssim8$ and preferentially removes the highest-redshift, reddest sources, rather than by the manifold selection itself (Section~\ref{sec:retention}). In addition, the broad-line search becomes reliable only above a moderate H$\alpha$ S/N (Section~\ref{sec:specmethods}). None of these limitations are
	fundamental to the approach and each points to a concrete way of strengthening it.
	
	Looking ahead, the framework extends naturally beyond photometry. 
	The same representation can ingest spectra and resolved imaging. 
	For instance, unsupervised methods have already been applied directly to JWST spectra 
	\citep[e.g.][]{Saxena25}, 
	and a closely related semi-supervised approach, anchoring a learned representation 
	with sparse labels (as we have here) has been used to classify AGNs in DESI spectra \citep{Alcolea26}. 
	The manifold can then propagate sparse, high-confidence labels through the learned space, and flag outliers and rare populations as they arise.

	\begin{acknowledgements}
		We are grateful to the organisers and participants of the hiking-workshop ``Galaxy and SMBH formation, growth, and co-evolution'' (Nepal, November 2025) for the fruitful discussions that inspired this work.
		FDE and RM acknowledge support by the Science and Technology Facilities Council (STFC), by the ERC through Advanced Grant 695671 ``QUENCH'', and by the UKRI Frontier Research grant RISEandFALL.
		RM also acknowledges funding from a research professorship from the Royal Society.
		GV acknowledges financial support by the Italian National Institute for Astrophysics (INAF) under the IAF - Astrophysics Fellowships in Italy grant CUP C59J21034720001 - ``AD MAJORA''.
		This work is based on observations made with the NASA/ESA/CSA James Webb Space Telescope. The data were obtained from the Mikulski Archive for Space Telescopes at the Space Telescope Science Institute, which is operated by the Association of Universities for Research in Astronomy, Inc., under NASA contract NAS 5-03127 for JWST. 
		These observations are associated with programs \#1180, \#1181, \#1210, \#1324, \#1345, \#1837, \#1895, \#2561, \#2756 and \#3990.
		We thank \citet{Merlin24} and the ASTRODEEP consortium for producing and releasing the ASTRODEEP-JWST photometric catalogue on which this analysis is based.
		Some of the data products presented herein were retrieved from the Dawn JWST Archive (DJA). DJA is an initiative of the Cosmic Dawn Center (DAWN), which is funded by the Danish National Research Foundation under grant DNRF140.
		This research made use of \textsc{numpy} \citep{Harris2020}, \textsc{scipy} \citep{Virtanen2020}, \textsc{scikit-learn} \citep{Pedregosa2011}, \textsc{matplotlib} \citep{Hunter2007}, \textsc{astropy} \citep{Astropy2022}, \textsc{pandas} \citep{pandasCollab} and \textsc{umap} \citep{McInnes18}.
	\end{acknowledgements}
	
	\bibliographystyle{bibtex/aa}
	\bibliography{references}
	
	\begin{appendix}
		\section{Feature-space classifier validation}
		\label{app:classifier}

\begin{figure}[t]\centering
			\includegraphics[width=\hsize]{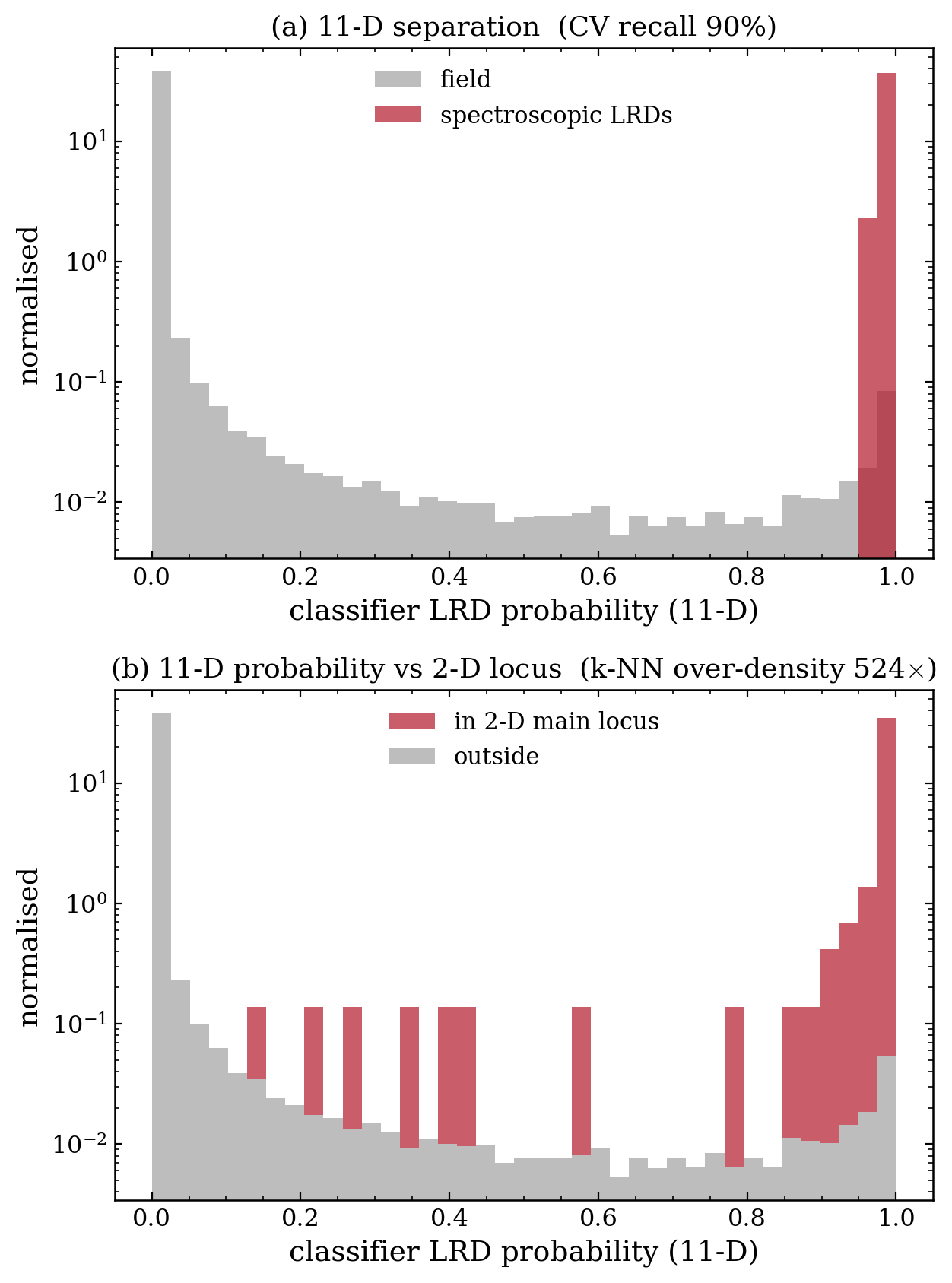}
			\caption{Feature-space validation of the LRD localisation. (a) Out-of-fold
				probabilities from a neural-network classifier trained on the original
				11D features, showing the separation between spectroscopic LRDs
				(red) and the field. (b) The same probabilities for sources inside and outside
				the 2D LRD region; the high-probability sources substantially
				overlap the manifold-defined region.}
			\label{fig:classifier}
		\end{figure}

		As an additional check that the LRD localisation is not a projection artefact, we
		trained a supervised classifier directly on the original feature vectors. Each
		source was represented by the same 11D vector used to construct the
		embedding: seven F356W-normalised log-colours, the log F356W reference flux, the
		stellarity, the log half-light radius, and $\log(1+z_{\rm phot})$. The features
		were robust-scaled before training.
		The classifier was a simple multilayer perceptron with two hidden layers and
		dropout regularisation, implemented in Keras/TensorFlow. We evaluated it with
		five-fold cross-validation: in each fold, the model was trained on four fifths of
		the spectroscopic anchors and the field, and predictions were recorded only for
		the held-out objects. The probabilities shown in Figure~\ref{fig:classifier} are
		therefore out-of-fold probabilities, so an anchor source does not contribute to
		the training set used to predict its own probability. This makes the test a
		validation of generalisation in feature space, rather than a restatement of the
		in-sample clustering.
		
		The resulting probability distributions separate the spectroscopic LRDs from the
		field and show substantial overlap between the high-probability sources and the
		2D LRD region. We use this classifier only as a diagnostic: the
		candidate selection adopted in the paper is still defined geometrically on the
		unsupervised manifold, anchored by the spectroscopic labels.

		\section{Follow-up candidates}
		\label{app:cand}
		
\begin{figure*}[h]\centering
			\includegraphics[width=\textwidth]{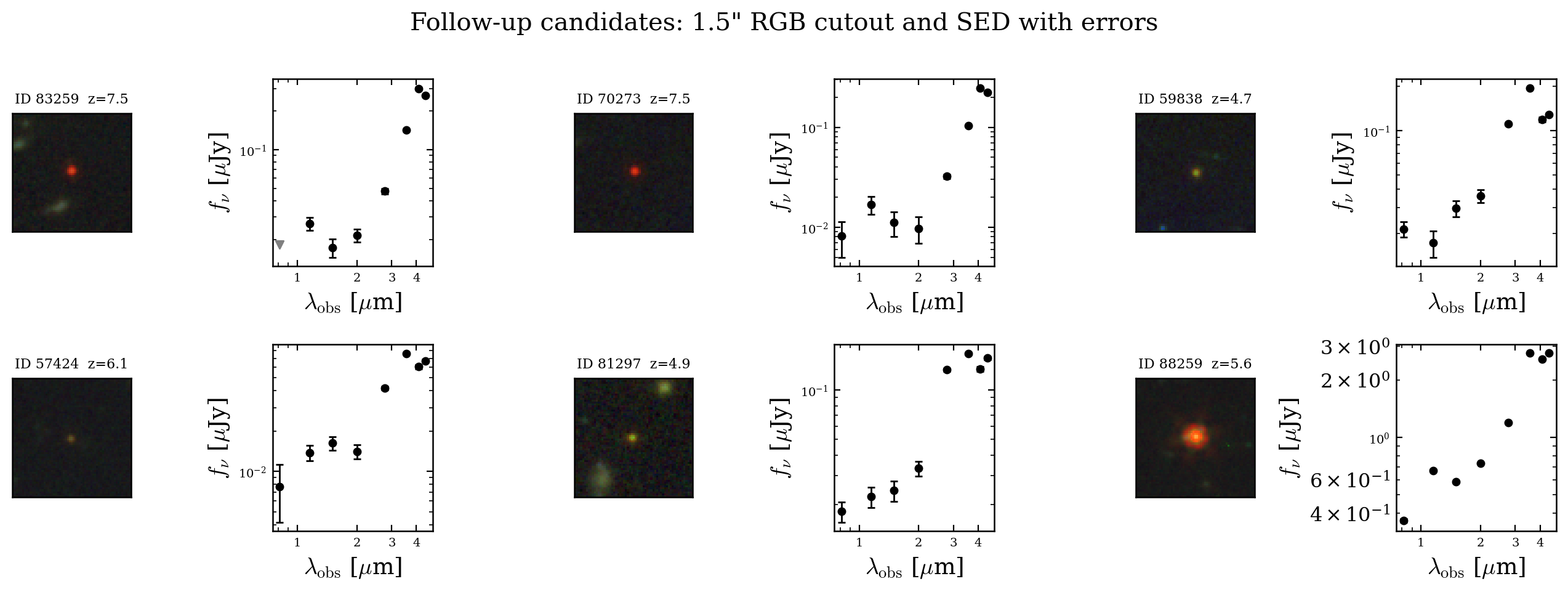}
			\caption{Follow-up candidates with no spectrum and no prior catalogue
				identification: for each, a $1.5''$ RGB cutout (left) and the observer-frame SED
				with uncertainties (right).}
			\label{fig:appcand}
		\end{figure*}

		Figure~\ref{fig:appcand} shows the most locus-central candidates from the main
		region (Section~\ref{sec:candidates}) that have no spectrum and no prior catalogue
		identification, ranked by Mahalanobis distance to the region centre. For each
		source we show a $1.5''\times1.5''$ RGB cutout (F150W/F277W/F444W) and the
		observer-frame SED with flux uncertainties (downwards triangles denote
		$3\sigma$ upper limits), so that the compactness and the red SED can be assessed
		directly. The complete ranked list, with coordinates, photometry and region
		membership, is provided as a machine-readable table
		\footnote{\url{https://github.com/micginolfi/compressedUniverse}}, to be released publicly upon acceptance.
		
		\section{Spectroscopic properties of the two loci}
		\label{app:populations}
\begin{figure*}[t]\centering
			\includegraphics[width=\textwidth]{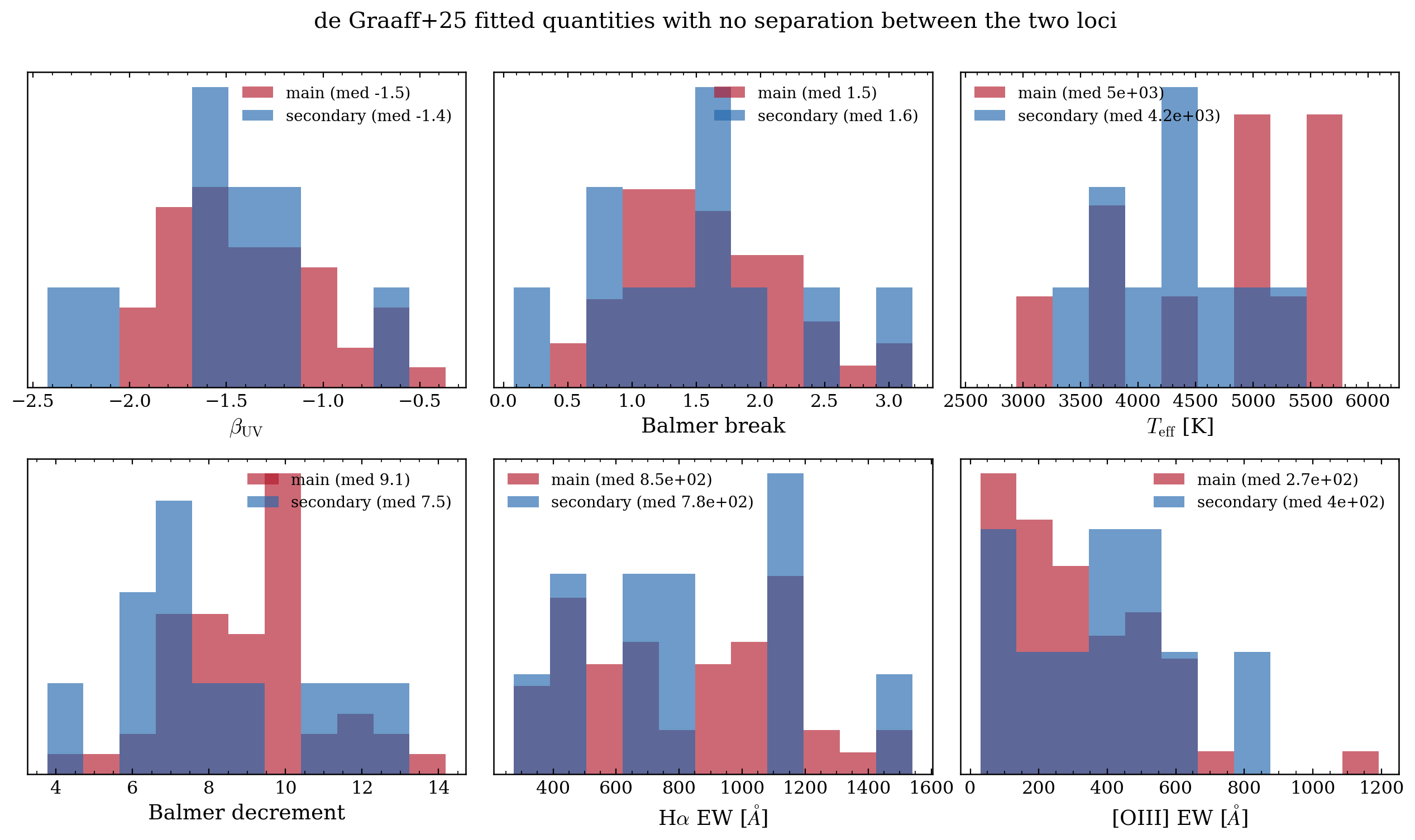}
			\caption{Fitted quantities from \citet{deGraaff25} that do not separate the two loci:
				rest-UV slope, $\beta_{\rm UV}$, Balmer-break strength, continuum
				temperature, $T_{\rm eff}$, Balmer decrement, and the H$\alpha$ and
				[O\,\textsc{iii}]\,$\lambda5007$ equivalent widths. Neither the main (red) nor the
				secondary (blue) island is distinguished by these, confirming that the two
				populations differ mainly in redshift and rest-UV luminosity
				(Figure~\ref{fig:twopop}).}
			\label{fig:populations}
		\end{figure*}

		Section~\ref{sec:populations} shows that the two anchor loci correspond to two
		populations differing mainly in redshift and rest-UV luminosity
		(Figure~\ref{fig:twopop}). For completeness, Figure~\ref{fig:populations} shows the
		remaining quantities from the public \citet{deGraaff25} spectral fits, matched to
		the anchors of each locus: the rest-UV slope, $\beta_{\rm UV}$, the
		Balmer-break strength, the continuum temperature, $T_{\rm eff}$, the Balmer
		decrement, and the H$\alpha$ and [O\,\textsc{iii}]\,$\lambda5007$ equivalent widths. 
		None of these separates the two loci, confirming that the distinction is driven by
		redshift and luminosity rather than by emission-line or continuum properties.

\begin{figure}[t]\centering
			\includegraphics[width=\hsize]{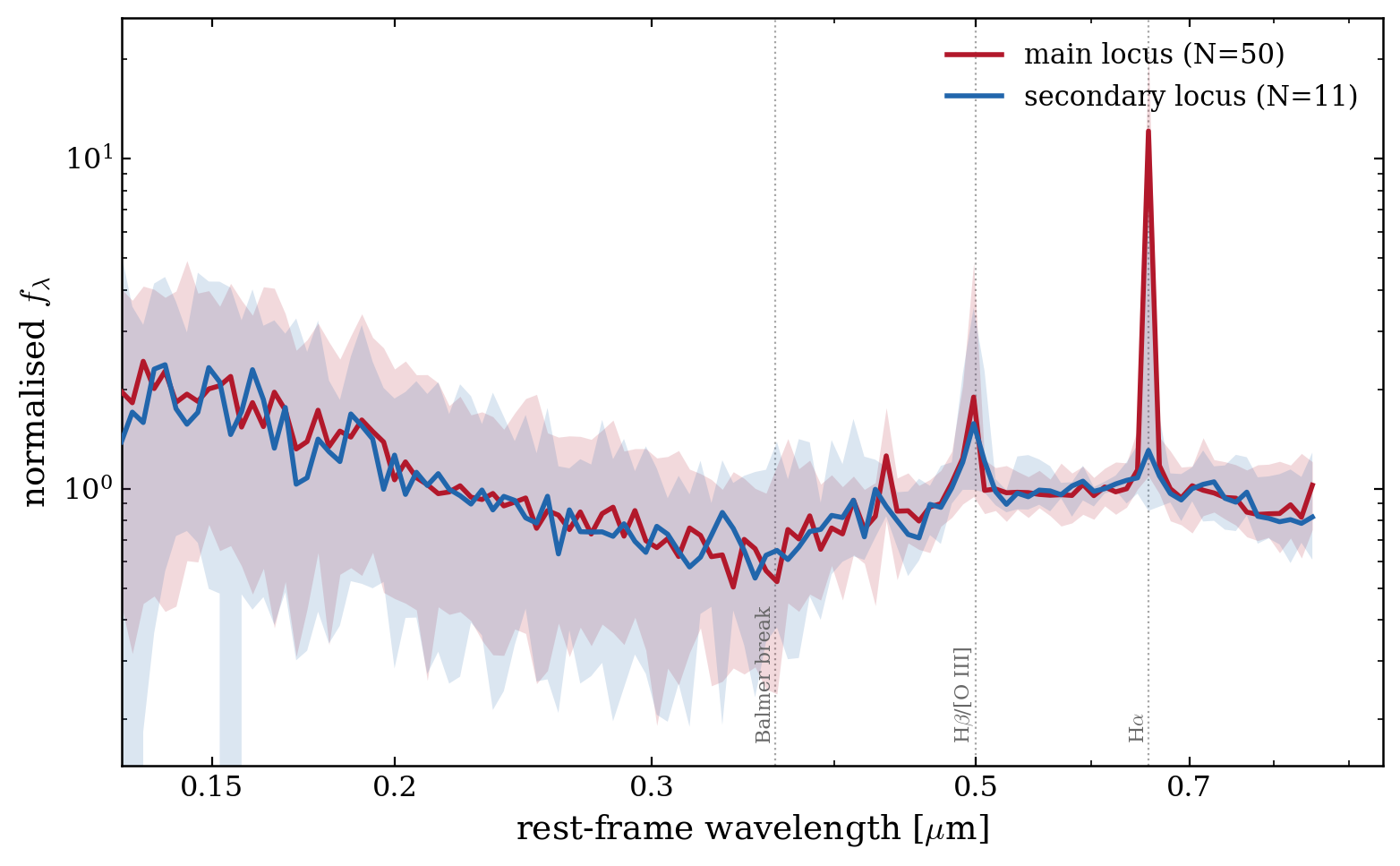}
			\caption{
				Rest-frame PRISM spectroscopic stacks of the \citet{deGraaff25} anchors in the main (red) and secondary (blue) loci, in $f_\lambda$ (median with 16--84th percentile band), each normalised to its median rest-optical (4500--6700~\AA) flux. Both loci show the same characteristic V-shape continuum and the same emission features (Balmer break, H$\beta$, [O\,\textsc{iii}]\,$\lambda5007$, H$\alpha$); the sharper lines in the main-locus stack (50 versus 11 spectra) reflect its larger sample, not a physical difference. The stacks reveal no spectral distinction between the loci beyond redshift and luminosity.
			}
			\label{fig:specstack}
		\end{figure}

		We further verify this directly from the spectra. Figure~\ref{fig:specstack} shows the stacked rest-frame PRISM spectra of the anchors in each locus. For each anchor we take its highest-S/N grade-3 PRISM spectrum from the DJA, shift it to the rest frame using its spectroscopic redshift, normalise it to its median flux over the rest-optical (4500--6700~\AA), resample onto a common rest-wavelength grid, and take the per-locus median with the 16--84th percentile range; we plot $f_\lambda$, in which the characteristic V-shape is most apparent. The two stacks (50 and 11 anchors with PRISM coverage in the main and secondary loci, respectively) are nearly identical in continuum shape and show the same emission features (H$\beta$, [O\,\textsc{iii}]\,$\lambda5007$ and H$\alpha$); the sharper lines in the main-locus stack reflect its larger sample rather than a physical difference. This makes explicit, directly from the spectra, the absence of any spectral distinction between the two loci beyond redshift and luminosity.
		
	\end{appendix}
	
\end{document}